\documentclass[a4paper,traditabstract,longauth]{aa}

\usepackage{graphicx}
\usepackage{txfonts}
\usepackage{natbib}
\usepackage{longtable,lscape}
\usepackage{amssymb}
\usepackage{mathrsfs}
\usepackage{amsfonts}
\usepackage{url}
\usepackage{color}
\usepackage[colorlinks, citecolor=blue]{hyperref}

\usepackage{fixltx2e}

\def\setsymbol#1#2{\expandafter\def\csname #1\endcsname{#2}}
\def\getsymbol#1{\csname #1\endcsname}

\def\Planck{\textit{Planck}}

\newbox\tablebox    \newdimen\tablewidth
\def\leaderfil{\leaders\hbox to 5pt{\hss.\hss}\hfil}
\def\endPlancktablewide{\tablewidth=\textwidth 
    $$\hss\copy\tablebox\hss$$
    \vskip-\lastskip\vskip -2pt}
\def\tablenote#1 #2\par{\begingroup \parindent=0.8em
    \abovedisplayshortskip=0pt\belowdisplayshortskip=0pt
    \noindent
    $$\hss\vbox{\hsize\tablewidth \hangindent=\parindent \hangafter=1 \noindent
    \hbox to \parindent{$^#1$\hss}\strut#2\strut\par}\hss$$
    \endgroup}
\def\doubleline{\vskip 3pt\hrule \vskip 1.5pt \hrule \vskip 5pt}

\def\L2{\ifmmode L_2\else $L_2$\fi}

\def\DeltaT{\ifmmode \Delta T\else $\Delta T$\fi}
\def\deltat{\ifmmode \Delta t\else $\Delta t$\fi}
\def\fknee{\ifmmode f_{\rm knee}\else $f_{\rm knee}$\fi}
\def\Fmax{\ifmmode F_{\rm max}\else $F_{\rm max}$\fi}
\def\solar{\ifmmode{\rm M}_{\mathord\odot}\else${\rm M}_{\mathord\odot}$\fi}
\def\Msolar{\ifmmode{\rm M}_{\mathord\odot}\else${\rm M}_{\mathord\odot}$\fi}
\def\Lsolar{\ifmmode{\rm L}_{\mathord\odot}\else${\rm L}_{\mathord\odot}$\fi}

\def\inv{\ifmmode^{-1}\else$^{-1}$\fi}
\def\mo{\ifmmode^{-1}\else$^{-1}$\fi}
\def\sup#1{\ifmmode ^{\rm #1}\else $^{\rm #1}$\fi}
\def\expo#1{\ifmmode \times 10^{#1}\else $\times 10^{#1}$\fi}
\def\,{\thinspace}
\def\lsim{\mathrel{\raise .4ex\hbox{\rlap{$<$}\lower 1.2ex\hbox{$\sim$}}}}
\def\gsim{\mathrel{\raise .4ex\hbox{\rlap{$>$}\lower 1.2ex\hbox{$\sim$}}}}

\def\simprop{\mathrel{\raise .4ex\hbox{\rlap{$\propto$}\lower 1.2ex\hbox{$\sim$}}}}
\def\deg{\ifmmode^\circ\else$^\circ$\fi}
\def\pdeg{\ifmmode $\setbox0=\hbox{$^{\circ}$}\rlap{\hskip.11\wd0 .}$^{\circ}
          \else \setbox0=\hbox{$^{\circ}$}\rlap{\hskip.11\wd0 .}$^{\circ}$\fi}
\def\arcs{\ifmmode {^{\scriptstyle\prime\prime}}
          \else $^{\scriptstyle\prime\prime}$\fi}
\def\arcm{\ifmmode {^{\scriptstyle\prime}}
          \else $^{\scriptstyle\prime}$\fi}
\newdimen\sa  \newdimen\sb
\def\parcs{\sa=.07em \sb=.03em
     \ifmmode \hbox{\rlap{.}}^{\scriptstyle\prime\kern -\sb\prime}\hbox{\kern -\sa}
     \else \rlap{.}$^{\scriptstyle\prime\kern -\sb\prime}$\kern -\sa\fi}
\def\parcm{\sa=.08em \sb=.03em
     \ifmmode \hbox{\rlap{.}\kern\sa}^{\scriptstyle\prime}\hbox{\kern-\sb}
     \else \rlap{.}\kern\sa$^{\scriptstyle\prime}$\kern-\sb\fi}
\def\ra[#1 #2 #3.#4]{#1\sup{h}#2\sup{m}#3\sup{s}\llap.#4}
\def\dec[#1 #2 #3.#4]{#1\deg#2\arcm#3\arcs\llap.#4}
\def\deco[#1 #2 #3]{#1\deg#2\arcm#3\arcs}
\def\rra[#1 #2]{#1\sup{h}#2\sup{m}}

\def\dots{\relax\ifmmode \ldots\else $\ldots$\fi}
\def\WHzsr{\ifmmode $W\,Hz\mo\,sr\mo$\else W\,Hz\mo\,sr\mo\fi}
\def\mHz{\ifmmode $\,mHz$\else \,mHz\fi}
\def\GHz{\ifmmode $\,GHz$\else \,GHz\fi}
\def\mKs{\ifmmode $\,mK\,s$^{1/2}\else \,mK\,s$^{1/2}$\fi}
\def\muKs{\ifmmode \,\mu$K\,s$^{1/2}\else \,$\mu$K\,s$^{1/2}$\fi}
\def\muKRJs{\ifmmode \,\mu$K$_{\rm RJ}$\,s$^{1/2}\else \,$\mu$K$_{\rm RJ}$\,s$^{1/2}$\fi}
\def\muKHz{\ifmmode \,\mu$K\,Hz$^{-1/2}\else \,$\mu$K\,Hz$^{-1/2}$\fi}
\def\MJysr{\ifmmode \,$MJy\,sr\mo$\else \,MJy\,sr\mo\fi}
\def\MJysrmK{\ifmmode \,$MJy\,sr\mo$\,mK$_{\rm CMB}\mo\else \,MJy\,sr\mo\,mK$_{\rm CMB}\mo$\fi}
\def\microns{\ifmmode \,\mu$m$\else \,$\mu$m\fi}

\def\muK{\ifmmode \,\mu$K$\else \,$\mu$\hbox{K}\fi}
\def\microK{\ifmmode \,\mu$K$\else \,$\mu$\hbox{K}\fi}
\def\muW{\ifmmode \,\mu$W$\else \,$\mu$\hbox{W}\fi}
\def\kms{\ifmmode $\,km\,s$^{-1}\else \,km\,s$^{-1}$\fi}
\def\kmsMpc{\ifmmode $\,\kms\,Mpc\mo$\else \,\kms\,Mpc\mo\fi}
\setsymbol{LFI:center:frequency:70GHz:units}{70.3\,GHz}
\setsymbol{LFI:center:frequency:44GHz:units}{44.1\,GHz}
\setsymbol{LFI:center:frequency:30GHz:units}{28.5\,GHz}

\setsymbol{LFI:center:frequency:70GHz}{70.3}
\setsymbol{LFI:center:frequency:44GHz}{44.1}
\setsymbol{LFI:center:frequency:30GHz}{28.5}

\setsymbol{LFI:center:frequency:LFI18:Rad:M:units}{71.7\GHz}
\setsymbol{LFI:center:frequency:LFI19:Rad:M:units}{67.5\GHz}
\setsymbol{LFI:center:frequency:LFI20:Rad:M:units}{69.2\GHz}
\setsymbol{LFI:center:frequency:LFI21:Rad:M:units}{70.4\GHz}
\setsymbol{LFI:center:frequency:LFI22:Rad:M:units}{71.5\GHz}
\setsymbol{LFI:center:frequency:LFI23:Rad:M:units}{70.8\GHz}
\setsymbol{LFI:center:frequency:LFI24:Rad:M:units}{44.4\GHz}
\setsymbol{LFI:center:frequency:LFI25:Rad:M:units}{44.0\GHz}
\setsymbol{LFI:center:frequency:LFI26:Rad:M:units}{43.9\GHz}
\setsymbol{LFI:center:frequency:LFI27:Rad:M:units}{28.3\GHz}
\setsymbol{LFI:center:frequency:LFI28:Rad:M:units}{28.8\GHz}
\setsymbol{LFI:center:frequency:LFI18:Rad:S:units}{70.1\GHz}
\setsymbol{LFI:center:frequency:LFI19:Rad:S:units}{69.6\GHz}
\setsymbol{LFI:center:frequency:LFI20:Rad:S:units}{69.5\GHz}
\setsymbol{LFI:center:frequency:LFI21:Rad:S:units}{69.5\GHz}
\setsymbol{LFI:center:frequency:LFI22:Rad:S:units}{72.8\GHz}
\setsymbol{LFI:center:frequency:LFI23:Rad:S:units}{71.3\GHz}
\setsymbol{LFI:center:frequency:LFI24:Rad:S:units}{44.1\GHz}
\setsymbol{LFI:center:frequency:LFI25:Rad:S:units}{44.1\GHz}
\setsymbol{LFI:center:frequency:LFI26:Rad:S:units}{44.1\GHz}
\setsymbol{LFI:center:frequency:LFI27:Rad:S:units}{28.5\GHz}
\setsymbol{LFI:center:frequency:LFI28:Rad:S:units}{28.2\GHz}

\setsymbol{LFI:center:frequency:LFI18:Rad:M}{71.7}
\setsymbol{LFI:center:frequency:LFI19:Rad:M}{67.5}
\setsymbol{LFI:center:frequency:LFI20:Rad:M}{69.2}
\setsymbol{LFI:center:frequency:LFI21:Rad:M}{70.4}
\setsymbol{LFI:center:frequency:LFI22:Rad:M}{71.5}
\setsymbol{LFI:center:frequency:LFI23:Rad:M}{70.8}
\setsymbol{LFI:center:frequency:LFI24:Rad:M}{44.4}
\setsymbol{LFI:center:frequency:LFI25:Rad:M}{44.0}
\setsymbol{LFI:center:frequency:LFI26:Rad:M}{43.9}
\setsymbol{LFI:center:frequency:LFI27:Rad:M}{28.3}
\setsymbol{LFI:center:frequency:LFI28:Rad:M}{28.8}
\setsymbol{LFI:center:frequency:LFI18:Rad:S}{70.1}
\setsymbol{LFI:center:frequency:LFI19:Rad:S}{69.6}
\setsymbol{LFI:center:frequency:LFI20:Rad:S}{69.5}
\setsymbol{LFI:center:frequency:LFI21:Rad:S}{69.5}
\setsymbol{LFI:center:frequency:LFI22:Rad:S}{72.8}
\setsymbol{LFI:center:frequency:LFI23:Rad:S}{71.3}
\setsymbol{LFI:center:frequency:LFI24:Rad:S}{44.1}
\setsymbol{LFI:center:frequency:LFI25:Rad:S}{44.1}
\setsymbol{LFI:center:frequency:LFI26:Rad:S}{44.1}
\setsymbol{LFI:center:frequency:LFI27:Rad:S}{28.5}
\setsymbol{LFI:center:frequency:LFI28:Rad:S}{28.2}

\setsymbol{LFI:white:noise:sensitivity:70GHz:units}{134.7\muKs}
\setsymbol{LFI:white:noise:sensitivity:44GHz:units}{164.7\muKs}
\setsymbol{LFI:white:noise:sensitivity:30GHz:units}{143.4\muKs}

\setsymbol{LFI:white:noise:sensitivity:70GHz}{134.7}
\setsymbol{LFI:white:noise:sensitivity:44GHz}{164.7}
\setsymbol{LFI:white:noise:sensitivity:30GHz}{143.4}

\setsymbol{LFI:white:noise:sensitivity:LFI18:Rad:M:units}{512.0\muKs}
\setsymbol{LFI:white:noise:sensitivity:LFI19:Rad:M:units}{581.4\muKs}
\setsymbol{LFI:white:noise:sensitivity:LFI20:Rad:M:units}{590.8\muKs}
\setsymbol{LFI:white:noise:sensitivity:LFI21:Rad:M:units}{455.2\muKs}
\setsymbol{LFI:white:noise:sensitivity:LFI22:Rad:M:units}{492.0\muKs}
\setsymbol{LFI:white:noise:sensitivity:LFI23:Rad:M:units}{507.7\muKs}
\setsymbol{LFI:white:noise:sensitivity:LFI24:Rad:M:units}{462.2\muKs}
\setsymbol{LFI:white:noise:sensitivity:LFI25:Rad:M:units}{413.6\muKs}
\setsymbol{LFI:white:noise:sensitivity:LFI26:Rad:M:units}{478.6\muKs}
\setsymbol{LFI:white:noise:sensitivity:LFI27:Rad:M:units}{277.7\muKs}
\setsymbol{LFI:white:noise:sensitivity:LFI28:Rad:M:units}{312.3\muKs}
\setsymbol{LFI:white:noise:sensitivity:LFI18:Rad:S:units}{465.7\muKs}
\setsymbol{LFI:white:noise:sensitivity:LFI19:Rad:S:units}{555.6\muKs}
\setsymbol{LFI:white:noise:sensitivity:LFI20:Rad:S:units}{623.2\muKs}
\setsymbol{LFI:white:noise:sensitivity:LFI21:Rad:S:units}{564.1\muKs}
\setsymbol{LFI:white:noise:sensitivity:LFI22:Rad:S:units}{534.4\muKs}
\setsymbol{LFI:white:noise:sensitivity:LFI23:Rad:S:units}{542.4\muKs}
\setsymbol{LFI:white:noise:sensitivity:LFI24:Rad:S:units}{399.2\muKs}
\setsymbol{LFI:white:noise:sensitivity:LFI25:Rad:S:units}{392.6\muKs}
\setsymbol{LFI:white:noise:sensitivity:LFI26:Rad:S:units}{418.6\muKs}
\setsymbol{LFI:white:noise:sensitivity:LFI27:Rad:S:units}{302.9\muKs}
\setsymbol{LFI:white:noise:sensitivity:LFI28:Rad:S:units}{285.3\muKs}

\setsymbol{LFI:white:noise:sensitivity:LFI18:Rad:M}{512.0}
\setsymbol{LFI:white:noise:sensitivity:LFI19:Rad:M}{581.4}
\setsymbol{LFI:white:noise:sensitivity:LFI20:Rad:M}{590.8}
\setsymbol{LFI:white:noise:sensitivity:LFI21:Rad:M}{455.2}
\setsymbol{LFI:white:noise:sensitivity:LFI22:Rad:M}{492.0}
\setsymbol{LFI:white:noise:sensitivity:LFI23:Rad:M}{507.7}
\setsymbol{LFI:white:noise:sensitivity:LFI24:Rad:M}{462.2}
\setsymbol{LFI:white:noise:sensitivity:LFI25:Rad:M}{413.6}
\setsymbol{LFI:white:noise:sensitivity:LFI26:Rad:M}{478.6}
\setsymbol{LFI:white:noise:sensitivity:LFI27:Rad:M}{277.7}
\setsymbol{LFI:white:noise:sensitivity:LFI28:Rad:M}{312.3}
\setsymbol{LFI:white:noise:sensitivity:LFI18:Rad:S}{465.7}
\setsymbol{LFI:white:noise:sensitivity:LFI19:Rad:S}{555.6}
\setsymbol{LFI:white:noise:sensitivity:LFI20:Rad:S}{623.2}
\setsymbol{LFI:white:noise:sensitivity:LFI21:Rad:S}{564.1}
\setsymbol{LFI:white:noise:sensitivity:LFI22:Rad:S}{534.4}
\setsymbol{LFI:white:noise:sensitivity:LFI23:Rad:S}{542.4}
\setsymbol{LFI:white:noise:sensitivity:LFI24:Rad:S}{399.2}
\setsymbol{LFI:white:noise:sensitivity:LFI25:Rad:S}{392.6}
\setsymbol{LFI:white:noise:sensitivity:LFI26:Rad:S}{418.6}
\setsymbol{LFI:white:noise:sensitivity:LFI27:Rad:S}{302.9}
\setsymbol{LFI:white:noise:sensitivity:LFI28:Rad:S}{285.3}

\setsymbol{LFI:knee:frequency:70GHz:units}{29.5\mHz}
\setsymbol{LFI:knee:frequency:44GHz:units}{56.2\mHz}
\setsymbol{LFI:knee:frequency:30GHz:units}{113.7\mHz}

\setsymbol{LFI:knee:frequency:70GHz}{29.5}
\setsymbol{LFI:knee:frequency:44GHz}{56.2}
\setsymbol{LFI:knee:frequency:30GHz}{113.7}

\setsymbol{LFI:knee:frequency:LFI18:Rad:M:units}{16.3\mHz}
\setsymbol{LFI:knee:frequency:LFI19:Rad:M:units}{15.1\mHz}
\setsymbol{LFI:knee:frequency:LFI20:Rad:M:units}{18.7\mHz}
\setsymbol{LFI:knee:frequency:LFI21:Rad:M:units}{37.2\mHz}
\setsymbol{LFI:knee:frequency:LFI22:Rad:M:units}{12.7\mHz}
\setsymbol{LFI:knee:frequency:LFI23:Rad:M:units}{34.6\mHz}
\setsymbol{LFI:knee:frequency:LFI24:Rad:M:units}{46.2\mHz}
\setsymbol{LFI:knee:frequency:LFI25:Rad:M:units}{24.9\mHz}
\setsymbol{LFI:knee:frequency:LFI26:Rad:M:units}{67.6\mHz}
\setsymbol{LFI:knee:frequency:LFI27:Rad:M:units}{187.4\mHz}
\setsymbol{LFI:knee:frequency:LFI28:Rad:M:units}{122.2\mHz}
\setsymbol{LFI:knee:frequency:LFI18:Rad:S:units}{17.7\mHz}
\setsymbol{LFI:knee:frequency:LFI19:Rad:S:units}{22.0\mHz}
\setsymbol{LFI:knee:frequency:LFI20:Rad:S:units}{8.7\mHz}
\setsymbol{LFI:knee:frequency:LFI21:Rad:S:units}{25.9\mHz}
\setsymbol{LFI:knee:frequency:LFI22:Rad:S:units}{15.8\mHz}
\setsymbol{LFI:knee:frequency:LFI23:Rad:S:units}{129.8\mHz}
\setsymbol{LFI:knee:frequency:LFI24:Rad:S:units}{100.9\mHz}
\setsymbol{LFI:knee:frequency:LFI25:Rad:S:units}{38.9\mHz}
\setsymbol{LFI:knee:frequency:LFI26:Rad:S:units}{58.9\mHz}
\setsymbol{LFI:knee:frequency:LFI27:Rad:S:units}{104.4\mHz}
\setsymbol{LFI:knee:frequency:LFI28:Rad:S:units}{40.7\mHz}

\setsymbol{LFI:knee:frequency:LFI18:Rad:M}{16.3}
\setsymbol{LFI:knee:frequency:LFI19:Rad:M}{15.1}
\setsymbol{LFI:knee:frequency:LFI20:Rad:M}{18.7}
\setsymbol{LFI:knee:frequency:LFI21:Rad:M}{37.2}
\setsymbol{LFI:knee:frequency:LFI22:Rad:M}{12.7}
\setsymbol{LFI:knee:frequency:LFI23:Rad:M}{34.6}
\setsymbol{LFI:knee:frequency:LFI24:Rad:M}{46.2}
\setsymbol{LFI:knee:frequency:LFI25:Rad:M}{24.9}
\setsymbol{LFI:knee:frequency:LFI26:Rad:M}{67.6}
\setsymbol{LFI:knee:frequency:LFI27:Rad:M}{187.4}
\setsymbol{LFI:knee:frequency:LFI28:Rad:M}{122.2}
\setsymbol{LFI:knee:frequency:LFI18:Rad:S}{17.7}
\setsymbol{LFI:knee:frequency:LFI19:Rad:S}{22.0}
\setsymbol{LFI:knee:frequency:LFI20:Rad:S}{8.7}
\setsymbol{LFI:knee:frequency:LFI21:Rad:S}{25.9}
\setsymbol{LFI:knee:frequency:LFI22:Rad:S}{15.8}
\setsymbol{LFI:knee:frequency:LFI23:Rad:S}{129.8}
\setsymbol{LFI:knee:frequency:LFI24:Rad:S}{100.9}
\setsymbol{LFI:knee:frequency:LFI25:Rad:S}{38.9}
\setsymbol{LFI:knee:frequency:LFI26:Rad:S}{58.9}
\setsymbol{LFI:knee:frequency:LFI27:Rad:S}{104.4}
\setsymbol{LFI:knee:frequency:LFI28:Rad:S}{40.7}

\setsymbol{LFI:slope:70GHz:units}{$-1.03$\mHz}
\setsymbol{LFI:slope:44GHz:units}{$-0.89$\mHz}
\setsymbol{LFI:slope:30GHz:units}{$-0.87$\mHz}

\setsymbol{LFI:slope:70GHz}{$-1.03$}
\setsymbol{LFI:slope:44GHz}{$-0.89$}
\setsymbol{LFI:slope:30GHz}{$-0.87$}

\setsymbol{LFI:slope:LFI18:Rad:M:units}{$-1.04$\mHz}
\setsymbol{LFI:slope:LFI19:Rad:M:units}{$-1.09$\mHz}
\setsymbol{LFI:slope:LFI20:Rad:M:units}{$-0.69$\mHz}
\setsymbol{LFI:slope:LFI21:Rad:M:units}{$-1.56$\mHz}
\setsymbol{LFI:slope:LFI22:Rad:M:units}{$-1.01$\mHz}
\setsymbol{LFI:slope:LFI23:Rad:M:units}{$-0.96$\mHz}
\setsymbol{LFI:slope:LFI24:Rad:M:units}{$-0.83$\mHz}
\setsymbol{LFI:slope:LFI25:Rad:M:units}{$-0.91$\mHz}
\setsymbol{LFI:slope:LFI26:Rad:M:units}{$-0.95$\mHz}
\setsymbol{LFI:slope:LFI27:Rad:M:units}{$-0.87$\mHz}
\setsymbol{LFI:slope:LFI28:Rad:M:units}{$-0.88$\mHz}
\setsymbol{LFI:slope:LFI18:Rad:S:units}{$-1.15$\mHz}
\setsymbol{LFI:slope:LFI19:Rad:S:units}{$-1.00$\mHz}
\setsymbol{LFI:slope:LFI20:Rad:S:units}{$-0.95$\mHz}
\setsymbol{LFI:slope:LFI21:Rad:S:units}{$-0.92$\mHz}
\setsymbol{LFI:slope:LFI22:Rad:S:units}{$-1.01$\mHz}
\setsymbol{LFI:slope:LFI23:Rad:S:units}{$-0.95$\mHz}
\setsymbol{LFI:slope:LFI24:Rad:S:units}{$-0.73$\mHz}
\setsymbol{LFI:slope:LFI25:Rad:S:units}{$-1.16$\mHz}
\setsymbol{LFI:slope:LFI26:Rad:S:units}{$-0.79$\mHz}
\setsymbol{LFI:slope:LFI27:Rad:S:units}{$-0.82$\mHz}
\setsymbol{LFI:slope:LFI28:Rad:S:units}{$-0.91$\mHz}

\setsymbol{LFI:slope:LFI18:Rad:M}{$-1.04$}
\setsymbol{LFI:slope:LFI19:Rad:M}{$-1.09$}
\setsymbol{LFI:slope:LFI20:Rad:M}{$-0.69$}
\setsymbol{LFI:slope:LFI21:Rad:M}{$-1.56$}
\setsymbol{LFI:slope:LFI22:Rad:M}{$-1.01$}
\setsymbol{LFI:slope:LFI23:Rad:M}{$-0.96$}
\setsymbol{LFI:slope:LFI24:Rad:M}{$-0.83$}
\setsymbol{LFI:slope:LFI25:Rad:M}{$-0.91$}
\setsymbol{LFI:slope:LFI26:Rad:M}{$-0.95$}
\setsymbol{LFI:slope:LFI27:Rad:M}{$-0.87$}
\setsymbol{LFI:slope:LFI28:Rad:M}{$-0.88$}
\setsymbol{LFI:slope:LFI18:Rad:S}{$-1.15$}
\setsymbol{LFI:slope:LFI19:Rad:S}{$-1.00$}
\setsymbol{LFI:slope:LFI20:Rad:S}{$-0.95$}
\setsymbol{LFI:slope:LFI21:Rad:S}{$-0.92$}
\setsymbol{LFI:slope:LFI22:Rad:S}{$-1.01$}
\setsymbol{LFI:slope:LFI23:Rad:S}{$-0.95$}
\setsymbol{LFI:slope:LFI24:Rad:S}{$-0.73$}
\setsymbol{LFI:slope:LFI25:Rad:S}{$-1.16$}
\setsymbol{LFI:slope:LFI26:Rad:S}{$-0.79$}
\setsymbol{LFI:slope:LFI27:Rad:S}{$-0.82$}
\setsymbol{LFI:slope:LFI28:Rad:S}{$-0.91$}

\setsymbol{LFI:FWHM:70GHz:units}{13\parcm01}
\setsymbol{LFI:FWHM:44GHz:units}{27\parcm92}
\setsymbol{LFI:FWHM:30GHz:units}{32\parcm65}

\setsymbol{LFI:FWHM:70GHz}{13.01}
\setsymbol{LFI:FWHM:44GHz}{27.92}
\setsymbol{LFI:FWHM:30GHz}{32.65}

\setsymbol{LFI:FWHM:LFI18:units}{13\parcm39}
\setsymbol{LFI:FWHM:LFI19:units}{13\parcm01}
\setsymbol{LFI:FWHM:LFI20:units}{12\parcm75}
\setsymbol{LFI:FWHM:LFI21:units}{12\parcm74}
\setsymbol{LFI:FWHM:LFI22:units}{12\parcm87}
\setsymbol{LFI:FWHM:LFI23:units}{13\parcm27}
\setsymbol{LFI:FWHM:LFI24:units}{22\parcm98}
\setsymbol{LFI:FWHM:LFI25:units}{30\parcm46}
\setsymbol{LFI:FWHM:LFI26:units}{30\parcm31}
\setsymbol{LFI:FWHM:LFI27:units}{32\parcm65}
\setsymbol{LFI:FWHM:LFI28:units}{32\parcm66}

\setsymbol{LFI:FWHM:LFI18}{13.39}
\setsymbol{LFI:FWHM:LFI19}{13.01}
\setsymbol{LFI:FWHM:LFI20}{12.75}
\setsymbol{LFI:FWHM:LFI21}{12.74}
\setsymbol{LFI:FWHM:LFI22}{12.87}
\setsymbol{LFI:FWHM:LFI23}{13.27}
\setsymbol{LFI:FWHM:LFI24}{22.98}
\setsymbol{LFI:FWHM:LFI25}{30.46}
\setsymbol{LFI:FWHM:LFI26}{30.31}
\setsymbol{LFI:FWHM:LFI27}{32.65}
\setsymbol{LFI:FWHM:LFI28}{32.66}

\setsymbol{LFI:FWHM:uncertainty:LFI18:units}{0.170\arcm}
\setsymbol{LFI:FWHM:uncertainty:LFI19:units}{0.174\arcm}
\setsymbol{LFI:FWHM:uncertainty:LFI20:units}{0.170\arcm}
\setsymbol{LFI:FWHM:uncertainty:LFI21:units}{0.156\arcm}
\setsymbol{LFI:FWHM:uncertainty:LFI22:units}{0.164\arcm}
\setsymbol{LFI:FWHM:uncertainty:LFI23:units}{0.171\arcm}
\setsymbol{LFI:FWHM:uncertainty:LFI24:units}{0.652\arcm}
\setsymbol{LFI:FWHM:uncertainty:LFI25:units}{1.075\arcm}
\setsymbol{LFI:FWHM:uncertainty:LFI26:units}{1.131\arcm}
\setsymbol{LFI:FWHM:uncertainty:LFI27:units}{1.266\arcm}
\setsymbol{LFI:FWHM:uncertainty:LFI28:units}{1.287\arcm}

\setsymbol{LFI:FWHM:uncertainty:LFI18}{0.170}
\setsymbol{LFI:FWHM:uncertainty:LFI19}{0.174}
\setsymbol{LFI:FWHM:uncertainty:LFI20}{0.170}
\setsymbol{LFI:FWHM:uncertainty:LFI21}{0.156}
\setsymbol{LFI:FWHM:uncertainty:LFI22}{0.164}
\setsymbol{LFI:FWHM:uncertainty:LFI23}{0.171}
\setsymbol{LFI:FWHM:uncertainty:LFI24}{0.652}
\setsymbol{LFI:FWHM:uncertainty:LFI25}{1.075}
\setsymbol{LFI:FWHM:uncertainty:LFI26}{1.131}
\setsymbol{LFI:FWHM:uncertainty:LFI27}{1.266}
\setsymbol{LFI:FWHM:uncertainty:LFI28}{1.287}

\setsymbol{HFI:center:frequency:100GHz:units}{100\,GHz}
\setsymbol{HFI:center:frequency:143GHz:units}{143\,GHz}
\setsymbol{HFI:center:frequency:217GHz:units}{217\,GHz}
\setsymbol{HFI:center:frequency:353GHz:units}{353\,GHz}
\setsymbol{HFI:center:frequency:545GHz:units}{545\,GHz}
\setsymbol{HFI:center:frequency:857GHz:units}{857\,GHz}

\setsymbol{HFI:center:frequency:100GHz}{100}
\setsymbol{HFI:center:frequency:143GHz}{143}
\setsymbol{HFI:center:frequency:217GHz}{217}
\setsymbol{HFI:center:frequency:353GHz}{353}
\setsymbol{HFI:center:frequency:545GHz}{545}
\setsymbol{HFI:center:frequency:857GHz}{857}

\setsymbol{HFI:Ndetectors:100GHz}{8}
\setsymbol{HFI:Ndetectors:143GHz}{11}
\setsymbol{HFI:Ndetectors:217GHz}{12}
\setsymbol{HFI:Ndetectors:353GHz}{12}
\setsymbol{HFI:Ndetectors:545GHz}{3}
\setsymbol{HFI:Ndetectors:857GHz}{4}

\setsymbol{HFI:FWHM:Maps:100GHz:units}{9\parcm88}
\setsymbol{HFI:FWHM:Maps:143GHz:units}{7\parcm18}
\setsymbol{HFI:FWHM:Maps:217GHz:units}{4\parcm87}
\setsymbol{HFI:FWHM:Maps:353GHz:units}{4\parcm65}
\setsymbol{HFI:FWHM:Maps:545GHz:units}{4\parcm72}
\setsymbol{HFI:FWHM:Maps:857GHz:units}{4\parcm39}
\setsymbol{HFI:FWHM:Maps:100GHz}{9.88}
\setsymbol{HFI:FWHM:Maps:143GHz}{7.18}
\setsymbol{HFI:FWHM:Maps:217GHz}{4.87}
\setsymbol{HFI:FWHM:Maps:353GHz}{4.65}
\setsymbol{HFI:FWHM:Maps:545GHz}{4.72}
\setsymbol{HFI:FWHM:Maps:857GHz}{4.39}

\setsymbol{HFI:beam:ellipticity:Maps:100GHz}{1.15}
\setsymbol{HFI:beam:ellipticity:Maps:143GHz}{1.01}
\setsymbol{HFI:beam:ellipticity:Maps:217GHz}{1.06}
\setsymbol{HFI:beam:ellipticity:Maps:353GHz}{1.05}
\setsymbol{HFI:beam:ellipticity:Maps:545GHz}{1.14}
\setsymbol{HFI:beam:ellipticity:Maps:857GHz}{1.19}

\setsymbol{HFI:FWHM:Mars:100GHz:units}{9\parcm37}
\setsymbol{HFI:FWHM:Mars:143GHz:units}{7\parcm04}
\setsymbol{HFI:FWHM:Mars:217GHz:units}{4\parcm68}
\setsymbol{HFI:FWHM:Mars:353GHz:units}{4\parcm43}
\setsymbol{HFI:FWHM:Mars:545GHz:units}{3\parcm80}
\setsymbol{HFI:FWHM:Mars:857GHz:units}{3\parcm67}

\setsymbol{HFI:FWHM:Mars:100GHz}{9.37}
\setsymbol{HFI:FWHM:Mars:143GHz}{7.04}
\setsymbol{HFI:FWHM:Mars:217GHz}{4.68}
\setsymbol{HFI:FWHM:Mars:353GHz}{4.43}
\setsymbol{HFI:FWHM:Mars:545GHz}{3.80}
\setsymbol{HFI:FWHM:Mars:857GHz}{3.67}

\setsymbol{HFI:beam:ellipticity:Mars:100GHz}{1.18}
\setsymbol{HFI:beam:ellipticity:Mars:143GHz}{1.03}
\setsymbol{HFI:beam:ellipticity:Mars:217GHz}{1.14}
\setsymbol{HFI:beam:ellipticity:Mars:353GHz}{1.09}
\setsymbol{HFI:beam:ellipticity:Mars:545GHz}{1.25}
\setsymbol{HFI:beam:ellipticity:Mars:857GHz}{1.03}

\setsymbol{HFI:CMB:relative:calibration:100GHz}{$\lsim 1\%$}
\setsymbol{HFI:CMB:relative:calibration:143GHz}{$\lsim 1\%$}
\setsymbol{HFI:CMB:relative:calibration:217GHz}{$\lsim 1\%$}
\setsymbol{HFI:CMB:relative:calibration:353GHz}{$\lsim 1\%$}
\setsymbol{HFI:CMB:relative:calibration:545GHz}{}
\setsymbol{HFI:CMB:relative:calibration:857GHz}{}

\setsymbol{HFI:CMB:absolute:calibration:100GHz}{$\lsim 2\%$}
\setsymbol{HFI:CMB:absolute:calibration:143GHz}{$\lsim 2\%$}
\setsymbol{HFI:CMB:absolute:calibration:217GHz}{$\lsim 2\%$}
\setsymbol{HFI:CMB:absolute:calibration:353GHz}{$\lsim 2\%$}
\setsymbol{HFI:CMB:absolute:calibration:545GHz}{}
\setsymbol{HFI:CMB:absolute:calibration:857GHz}{}

\setsymbol{HFI:FIRAS:gain:calibration:accuracy:statistical:100GHz}{}
\setsymbol{HFI:FIRAS:gain:calibration:accuracy:statistical:143GHz}{}
\setsymbol{HFI:FIRAS:gain:calibration:accuracy:statistical:217GHz}{}
\setsymbol{HFI:FIRAS:gain:calibration:accuracy:statistical:353GHz}{2.5\%}
\setsymbol{HFI:FIRAS:gain:calibration:accuracy:statistical:545GHz}{1\%}
\setsymbol{HFI:FIRAS:gain:calibration:accuracy:statistical:857GHz}{0.5\%}

\setsymbol{HFI:FIRAS:gain:calibration:accuracy:systematic:100GHz}{}
\setsymbol{HFI:FIRAS:gain:calibration:accuracy:systematic:143GHz}{}
\setsymbol{HFI:FIRAS:gain:calibration:accuracy:systematic:217GHz}{}
\setsymbol{HFI:FIRAS:gain:calibration:accuracy:systematic:353GHz}{}
\setsymbol{HFI:FIRAS:gain:calibration:accuracy:systematic:545GHz}{7\%}
\setsymbol{HFI:FIRAS:gain:calibration:accuracy:systematic:857GHz}{7\%}

\setsymbol{HFI:FIRAS:zero:point:accuracy:100GHz:units}{0.8\MJysr}
\setsymbol{HFI:FIRAS:zero:point:accuracy:143GHz:units}{}
\setsymbol{HFI:FIRAS:zero:point:accuracy:217GHz:units}{}
\setsymbol{HFI:FIRAS:zero:point:accuracy:353GHz:units}{1.4\MJysr}
\setsymbol{HFI:FIRAS:zero:point:accuracy:545GHz:units}{2.2\MJysr}
\setsymbol{HFI:FIRAS:zero:point:accuracy:857GHz:units}{1.7\MJysr}

\setsymbol{HFI:FIRAS:zero:point:accuracy:100GHz}{0.8}
\setsymbol{HFI:FIRAS:zero:point:accuracy:143GHz}{}
\setsymbol{HFI:FIRAS:zero:point:accuracy:217GHz}{}
\setsymbol{HFI:FIRAS:zero:point:accuracy:353GHz}{1.4}
\setsymbol{HFI:FIRAS:zero:point:accuracy:545GHz}{2.2}
\setsymbol{HFI:FIRAS:zero:point:accuracy:857GHz}{1.7}

\setsymbol{HFI:unit:conversion:100GHz:units}{0.2415\MJysrmK}
\setsymbol{HFI:unit:conversion:143GHz:units}{0.3694\MJysrmK}
\setsymbol{HFI:unit:conversion:217GHz:units}{0.4811\MJysrmK}
\setsymbol{HFI:unit:conversion:353GHz:units}{0.2883\MJysrmK}
\setsymbol{HFI:unit:conversion:545GHz:units}{0.05826\MJysrmK}
\setsymbol{HFI:unit:conversion:857GHz:units}{0.002238\MJysrmK}

\setsymbol{HFI:unit:conversion:100GHz}{0.2415}
\setsymbol{HFI:unit:conversion:143GHz}{0.3694}
\setsymbol{HFI:unit:conversion:217GHz}{0.4811}
\setsymbol{HFI:unit:conversion:353GHz}{0.2883}
\setsymbol{HFI:unit:conversion:545GHz}{0.05826}
\setsymbol{HFI:unit:conversion:857GHz}{0.002238}

\setsymbol{HFI:colour:correction:alpha=-2:V1.01:100GHz}{0.9893}
\setsymbol{HFI:colour:correction:alpha=-2:V1.01:143GHz}{0.9759}
\setsymbol{HFI:colour:correction:alpha=-2:V1.01:217GHz}{1.0007}
\setsymbol{HFI:colour:correction:alpha=-2:V1.01:353GHz}{1.0028}
\setsymbol{HFI:colour:correction:alpha=-2:V1.01:545GHz}{1.0019}
\setsymbol{HFI:colour:correction:alpha=-2:V1.01:857GHz}{0.9889}

\setsymbol{HFI:colour:correction:alpha=0:V1.01:100GHz}{1.0008}
\setsymbol{HFI:colour:correction:alpha=0:V1.01:143GHz}{1.0148}
\setsymbol{HFI:colour:correction:alpha=0:V1.01:217GHz}{0.9909}
\setsymbol{HFI:colour:correction:alpha=0:V1.01:353GHz}{0.9888}
\setsymbol{HFI:colour:correction:alpha=0:V1.01:545GHz}{0.9878}
\setsymbol{HFI:colour:correction:alpha=0:V1.01:857GHz}{1.0014}

\begin{document}

\title{\Planck\ intermediate results. XXVI. Optical identification and
  redshifts of \Planck\ clusters with the RTT150 telescope}


\author{\small
Planck Collaboration:
P.~A.~R.~Ade\inst{74}
\and
N.~Aghanim\inst{51}
\and
M.~Arnaud\inst{63}
\and
M.~Ashdown\inst{60, 6}
\and
J.~Aumont\inst{51}
\and
C.~Baccigalupi\inst{73}
\and
A.~J.~Banday\inst{81, 10}
\and
R.~B.~Barreiro\inst{57}
\and
R.~Barrena\inst{56}
\and
N.~Bartolo\inst{28}
% ...
}
\institute{\small
APC, AstroParticule et Cosmologie, Universit\'{e} Paris Diderot, CNRS/IN2P3, CEA/lrfu, Observatoire de Paris, Sorbonne Paris Cit\'{e}, 10, rue Alice Domon et L\'{e}onie Duquet, 75205 Paris Cedex 13, France\\
\and
Academy of Sciences of Tatarstan, Bauman Str., 20, Kazan, 420111, Republic of Tatarstan, Russia\\
\and
African Institute for Mathematical Sciences, 6-8 Melrose Road, Muizenberg, Cape Town, South Africa\\
\and
Agenzia Spaziale Italiana Science Data Center, Via del Politecnico snc, 00133, Roma, Italy\\
\and
Agenzia Spaziale Italiana, Viale Liegi 26, Roma, Italy\\
\and
Astrophysics Group, Cavendish Laboratory, University of Cambridge, J J Thomson Avenue, Cambridge CB3 0HE, U.K.\\
\and
Astrophysics \& Cosmology Research Unit, School of Mathematics, Statistics \& Computer Science, University of KwaZulu-Natal, Westville Campus, Private Bag X54001, Durban 4000, South Africa\\
\and
Atacama Large Millimeter/submillimeter Array, ALMA Santiago Central Offices, Alonso de Cordova 3107, Vitacura, Casilla 763 0355, Santiago, Chile\\
\and
CITA, University of Toronto, 60 St. George St., Toronto, ON M5S 3H8, Canada\\
\and
CNRS, IRAP, 9 Av. colonel Roche, BP 44346, F-31028 Toulouse cedex 4, France\\
\and
California Institute of Technology, Pasadena, California, U.S.A.\\
\and
Centro de Estudios de F\'{i}sica del Cosmos de Arag\'{o}n (CEFCA), Plaza San Juan, 1, planta 2, E-44001, Teruel, Spain\\
\and
Computational Cosmology Center, Lawrence Berkeley National Laboratory, Berkeley, California, U.S.A.\\
\and
DSM/Irfu/SPP, CEA-Saclay, F-91191 Gif-sur-Yvette Cedex, France\\
\and
DTU Space, National Space Institute, Technical University of Denmark, Elektrovej 327, DK-2800 Kgs. Lyngby, Denmark\\
\and
D\'{e}partement de Physique Th\'{e}orique, Universit\'{e} de Gen\`{e}ve, 24, Quai E. Ansermet,1211 Gen\`{e}ve 4, Switzerland\\
\and
Departamento de F\'{\i}sica, Universidad de Oviedo, Avda. Calvo Sotelo s/n, Oviedo, Spain\\
\and
Department of Astronomy and Geodesy, Kazan Federal University,  Kremlevskaya Str., 18, Kazan, 420008, Russia\\
\and
Department of Astrophysics/IMAPP, Radboud University Nijmegen, P.O. Box 9010, 6500 GL Nijmegen, The Netherlands\\
\and
Department of Physics \& Astronomy, University of British Columbia, 6224 Agricultural Road, Vancouver, British Columbia, Canada\\
\and
Department of Physics and Astronomy, Dana and David Dornsife College of Letter, Arts and Sciences, University of Southern California, Los Angeles, CA 90089, U.S.A.\\
\and
Department of Physics and Astronomy, University College London, London WC1E 6BT, U.K.\\
\and
Department of Physics, Florida State University, Keen Physics Building, 77 Chieftan Way, Tallahassee, Florida, U.S.A.\\
\and
Department of Physics, Gustaf H\"{a}llstr\"{o}min katu 2a, University of Helsinki, Helsinki, Finland\\
\and
Department of Physics, Princeton University, Princeton, New Jersey, U.S.A.\\
\and
Department of Physics, University of California, Santa Barbara, California, U.S.A.\\
\and
Department of Physics, University of Illinois at Urbana-Champaign, 1110 West Green Street, Urbana, Illinois, U.S.A.\\
\and
Dipartimento di Fisica e Astronomia G. Galilei, Universit\`{a} degli Studi di Padova, via Marzolo 8, 35131 Padova, Italy\\
\and
Dipartimento di Fisica e Scienze della Terra, Universit\`{a} di Ferrara, Via Saragat 1, 44122 Ferrara, Italy\\
\and
Dipartimento di Fisica, Universit\`{a} La Sapienza, P. le A. Moro 2, Roma, Italy\\
\and
Dipartimento di Fisica, Universit\`{a} degli Studi di Milano, Via Celoria, 16, Milano, Italy\\
\and
Dipartimento di Fisica, Universit\`{a} di Roma Tor Vergata, Via della Ricerca Scientifica, 1, Roma, Italy\\
\and
Discovery Center, Niels Bohr Institute, Blegdamsvej 17, Copenhagen, Denmark\\
\and
Dpto. Astrof\'{i}sica, Universidad de La Laguna (ULL), E-38206 La Laguna, Tenerife, Spain\\
\and
European Southern Observatory, ESO Vitacura, Alonso de Cordova 3107, Vitacura, Casilla 19001, Santiago, Chile\\
\and
European Space Agency, ESAC, Planck Science Office, Camino bajo del Castillo, s/n, Urbanizaci\'{o}n Villafranca del Castillo, Villanueva de la Ca\~{n}ada, Madrid, Spain\\
\and
European Space Agency, ESTEC, Keplerlaan 1, 2201 AZ Noordwijk, The Netherlands\\
\and
Facolt\`{a} di Ingegneria, Universit\`{a} degli Studi e-Campus, Via Isimbardi 10, Novedrate (CO), 22060, Italy\\
\and
HGSFP and University of Heidelberg, Theoretical Physics Department, Philosophenweg 16, 69120, Heidelberg, Germany\\
\and
Helsinki Institute of Physics, Gustaf H\"{a}llstr\"{o}min katu 2, University of Helsinki, Helsinki, Finland\\
\and
INAF - Osservatorio Astronomico di Padova, Vicolo dell'Osservatorio 5, Padova, Italy\\
\and
INAF - Osservatorio Astronomico di Roma, via di Frascati 33, Monte Porzio Catone, Italy\\
\and
INAF - Osservatorio Astronomico di Trieste, Via G.B. Tiepolo 11, Trieste, Italy\\
\and
INAF/IASF Bologna, Via Gobetti 101, Bologna, Italy\\
\and
INAF/IASF Milano, Via E. Bassini 15, Milano, Italy\\
\and
INFN, Sezione di Bologna, Via Irnerio 46, I-40126, Bologna, Italy\\
\and
INFN, Sezione di Roma 1, Universit\`{a} di Roma Sapienza, Piazzale Aldo Moro 2, 00185, Roma, Italy\\
\and
Imperial College London, Astrophysics group, Blackett Laboratory, Prince Consort Road, London, SW7 2AZ, U.K.\\
\and
Infrared Processing and Analysis Center, California Institute of Technology, Pasadena, CA 91125, U.S.A.\\
\and
Institut Universitaire de France, 103, bd Saint-Michel, 75005, Paris, France\\
\and
Institut d'Astrophysique Spatiale, CNRS (UMR8617) Universit\'{e} Paris-Sud 11, B\^{a}timent 121, Orsay, France\\
\and
Institut d'Astrophysique de Paris, CNRS (UMR7095), 98 bis Boulevard Arago, F-75014, Paris, France\\
\and
Institute for Space Sciences, Bucharest-Magurale, Romania\\
\and
Institute of Astronomy, University of Cambridge, Madingley Road, Cambridge CB3 0HA, U.K.\\
\and
Institute of Theoretical Astrophysics, University of Oslo, Blindern, Oslo, Norway\\
\and
Instituto de Astrof\'{\i}sica de Canarias, C/V\'{\i}a L\'{a}ctea s/n, La Laguna, Tenerife, Spain\\
\and
Instituto de F\'{\i}sica de Cantabria (CSIC-Universidad de Cantabria), Avda. de los Castros s/n, Santander, Spain\\
\and
Jet Propulsion Laboratory, California Institute of Technology, 4800 Oak Grove Drive, Pasadena, California, U.S.A.\\
\and
Jodrell Bank Centre for Astrophysics, Alan Turing Building, School of Physics and Astronomy, The University of Manchester, Oxford Road, Manchester, M13 9PL, U.K.\\
\and
Kavli Institute for Cosmology Cambridge, Madingley Road, Cambridge, CB3 0HA, U.K.\\
\and
LAL, Universit\'{e} Paris-Sud, CNRS/IN2P3, Orsay, France\\
\and
LERMA, CNRS, Observatoire de Paris, 61 Avenue de l'Observatoire, Paris, France\\
\and
Laboratoire AIM, IRFU/Service d'Astrophysique - CEA/DSM - CNRS - Universit\'{e} Paris Diderot, B\^{a}t. 709, CEA-Saclay, F-91191 Gif-sur-Yvette Cedex, France\\
\and
Laboratoire de Physique Subatomique et de Cosmologie, Universit\'{e} Joseph Fourier Grenoble I, CNRS/IN2P3, Institut National Polytechnique de Grenoble, 53 rue des Martyrs, 38026 Grenoble cedex, France\\
\and
Laboratoire de Physique Th\'{e}orique, Universit\'{e} Paris-Sud 11 \& CNRS, B\^{a}timent 210, 91405 Orsay, France\\
\and
Lawrence Berkeley National Laboratory, Berkeley, California, U.S.A.\\
\and
Max-Planck-Institut f\"{u}r Astrophysik, Karl-Schwarzschild-Str. 1, 85741 Garching, Germany\\
\and
Max-Planck-Institut f\"{u}r Extraterrestrische Physik, Giessenbachstra{\ss}e, 85748 Garching, Germany\\
\and
McGill Physics, Ernest Rutherford Physics Building, McGill University, 3600 rue University, Montr\'{e}al, QC, H3A 2T8, Canada\\
\and
Moscow Institute of Physics and Technology, Dolgoprudny, Institutsky per., 9, 141700, Russia\\
\and
National University of Ireland, Department of Experimental Physics, Maynooth, Co. Kildare, Ireland\\
\and
Niels Bohr Institute, Blegdamsvej 17, Copenhagen, Denmark\\
\and
SISSA, Astrophysics Sector, via Bonomea 265, 34136, Trieste, Italy\\
\and
School of Physics and Astronomy, Cardiff University, Queens Buildings, The Parade, Cardiff, CF24 3AA, U.K.\\
\and
Space Research Institute (IKI), Russian Academy of Sciences, Profsoyuznaya Str, 84/32, Moscow, 117997, Russia\\
\and
Space Sciences Laboratory, University of California, Berkeley, California, U.S.A.\\
\and
Special Astrophysical Observatory, Russian Academy of Sciences, Nizhnij Arkhyz, Zelenchukskiy region, Karachai-Cherkessian Republic, 369167, Russia\\
\and
Sub-Department of Astrophysics, University of Oxford, Keble Road, Oxford OX1 3RH, U.K.\\
\and
T\"{U}B\.{I}TAK National Observatory, Akdeniz University Campus, 07058, Antalya, Turkey\\
\and
UPMC Univ Paris 06, UMR7095, 98 bis Boulevard Arago, F-75014, Paris, France\\
\and
Universit\'{e} de Toulouse, UPS-OMP, IRAP, F-31028 Toulouse cedex 4, France\\
% ...
}


\abstract{ We present the results of approximately three years of
  observations of \Planck\ Sunyaev-Zeldovich (SZ) sources with the
  Russian-Turkish 1.5-m telescope (RTT150), as a part of the optical
  follow-up programme undertaken by the \Planck\ collaboration. During
  this time period approximately 20\,\% of all dark and grey clear
  time available at the telescope was devoted to observations of {\it
    Planck} objects. Some observations of distant clusters were also
  done at the 6-m Bolshoy Telescope Azimutal'ny (BTA) of the Special
  Astrophysical Observatory of the Russian Academy of Sciences. In
  total, deep, direct images of more than one hundred fields were
  obtained in multiple filters. We identified 47 previously unknown
  galaxy clusters, 41 of which are included in the \Planck\ catalogue
  of SZ sources. The redshifts of 65 
  \Planck\ clusters were measured spectroscopically and 14 more were
  measured photometrically. We discuss the details of cluster optical
  identifications and redshift measurements. We also present new
  spectroscopic redhifts for 39 
  \Planck\ clusters that were not included in the \Planck\ SZ source
  catalogue and are published here for the first time.}


  \authorrunning{Planck Collaboration}

  \titlerunning{\Planck\ SZ source catalogue: Optical identification
    using RTT150 telescope}

\maketitle

\section{Introduction}

The {\it Planck} all-sky survey is the first survey in which a
significant number of galaxy clusters have been detected by means of
the Sunyaev-Zeldovich (SZ) effect \citep{sz72} over all the entire
extragalactic sky \citep{PC11a,PSZcat13}. Since the SZ signal does not
suffer from cosmological dimming and is approximately proportional to
cluster mass, the {\it Planck} SZ galaxy cluster survey contains the
most massive clusters in the Universe, and is therefore of unique
importance for cluster and cosmological studies.

While blind SZ cluster detection with \Planck\ is robust \citep[see,
e.g.,][]{PSZcat13}, further observations of newly detected candidate
clusters at other wavelengths are still required. The \Planck\
collaboration has undertaken an extensive follow-up programme to confirm 
\Planck\ cluster candidates taken from intermediate versions of the \Planck\ SZ
catalogue and to measure their redshifts (e.g., \citealt{planck2011-5.1b}; 
\citealt{planck2012-I}; \citealt{planck2012-IV}), using the 
European Northern and Southern observatories (ENO and ESO) and other
telescopes. The strategy of this follow-up programme is detailed in
\citet{PC11a,PSZcat13}.

In this paper, we describe observations with the Russian-Turkish 1.5-m
telescope (RTT150\footnote{http://hea.iki.rssi.ru/rtt150/en/}).  Over
three years, approximately 20$\,$\% of all clear dark and grey time
available at the telescope was used to observe {\it Planck} SZ
sources.  Cluster identification procedures are based on those
developed for the 400\,deg$^2$ X-ray galaxy cluster survey
\citep[400d,][]{400d}, and for an earlier 160\,deg$^2$ survey
\citep{160d,mullis03}. As was shown with the last two surveys and will
be demonstrated below, with 1.5-m class telescopes clusters can be
identified at redshifts up to $z\approx1$, and redshifts can be
measured spectroscopically for clusters at approximately
$z<0.4$. Therefore, data taken with such telescopes are sufficient to
provide optical identifications and redshift measurements for a large
fraction of galaxy clusters detected with \Planck.

A large fraction of the cluster identifications and redshift
measurements presented in this paper were included in the recently
published \Planck\ SZ source catalogue, \citep{PSZcat13}. 
This companion article presents the details of optical
identifications, results of more recent observations at RTT150, and
the optical identifications for some \Planck\ cluster candidates below
the S/N = 4.5 limit of the \Planck\ catalogue.


The paper is organised as follows. Section~\ref{sec:rtt150} describes
the RTT150 telescope and the \Planck\ cluster observing programme
carried out on it.  Sections~\ref{sec:plcl} and \ref{sec:opt} review
the procedures used for cluster selection and optical identification,
and discuss the observations themselves.  Finally,
Sects.~\ref{sec:obs} and \ref{sec:res} describe in detail the results
of the observations, give examples of cluster identifications, and
discuss both individual objects and probable false SZ sources
identified in our programme.

\section{The RTT150 telescope}
\label{sec:rtt150}

The RTT150 optics are of high quality \citep{2001AstL...27..398A} and
the telescope site (T\"{U}B\.ITAK National Observatory, Bakyrlytepe
mountain, altitude 2550\,m, location
$2^\mathrm{h}01^\mathrm{m}20^\mathrm{s}$~E, $36^\circ49'30''$~N) has
good astronomical weather.  We used the TFOSC instrument
(T\"{U}B\.ITAK Faint Object Spectrograph and Camera), a focal-reducer
type spectrograph and camera built at Copenhagen University
Observatory. This instrument is similar to ALFOSC at the Nordic
Optical Telescope (NOT), also used in the \Planck\ follow-up programme
of clusters, and to other instruments of this series.

TFOSC is equipped with Bessel, Sloan Digital Sky Survey
(SDSS), and other filter-sets. It allows a quick switch between
direct imaging and spectroscopic modes, which increases the efficiency
of the instrument.  The size of the TFOSC field of view in direct-imaging
mode is $13\parcm3\times13\parcm3$, with a $0\parcs39$ pixel
scale. In spectroscopic mode, the instrument allows us to obtain low
and medium resolution ($500\lsim R\lsim 5000$) long-slit spectra.

The \Planck\ cluster follow-up programme was started at the RTT150
telescope in the summer of 2011. We present here the results of
observations obtained through the spring of 2014. During this period
approximately 60 \emph{clear} dark and grey nights were used.  As
previously stated, this corresponds to approximately 20\% of total
amount of dark and grey clear time available at the telescope. This
observing time was provided by the Kazan Federal University (KFU) and
the Space Research Institute (IKI), operators of the RTT150 from the
Russian side.


Additional observations of clusters at high redshift were made with
the 6-m Bolshoy Teleskop Azimutal'nyi (BTA) of the Special
Astrophysical Observatory of the Russian Academy of Sciences, using
the SCORPIO spectrograph \citep{scoprpio}, which is similar in layout
and capabilities to the TFOSC instrument at RTT150 and is optimized
for spectroscopic observations of faint objects.  We used
approximately 32~hours of clear weather at the BTA for the \Planck\
cluster follow-up programme.

\section{\Planck\ cluster selection}
\label{sec:plcl}

The \Planck\ catalogue of SZ sources \cite[PSZ1 hereafter,
][]{PSZcat13} consists of 1227 sources detected through their SZ
effect above S/N=4.5 in the \Planck\ frequency maps. The catalogue
contains a large fraction of newly confirmed or previously known
galaxy clusters but it also contains un-confirmed cluster candidates.
The procedures used for cluster selection and identification in the
{\it Planck} cluster survey are discussed in detail in
\cite{PSZcat13}.  The main steps are as follows.

Cluster candidates are detected blindly using three different
algorithms.  The quality of an individual detection is estimated from
the significance of the SZ signal, the agreement between the three
algorithms, and the frequency spectrum of the cluster candidate.

Cluster candidates are cross-correlated with optical, X-ray, and other
SZ catalogues and samples.  Detection of X-ray emission from the same
hot intracluster gas that {\it Planck} detects through the SZ effect
provides definitive confirmation.  Cluster candidates are therefore
checked for counterparts in the {\it ROSAT} All Sky Survey
\citep[RASS,][]{rassbr,rassfaint}.  Since \Planck\ detects the most
massive and the most X-ray-luminous clusters, \Planck\ candidates
should be detectable in RASS up to $z\approx0.3$--$0.4$.  Candidates
are also checked for counterparts in the Sloan Digital Sky Survey
\citep[SDSS, DR8,][]{2011ApJS..193...29A} and the {\it WISE} all-sky
survey \citep{2010AJ....140.1868W}.


Candidates without confirmation from these various steps were sent to
observing facilities for follow-up observations to confirm them as
clusters and to measure their redshifts.  In particular,
\Planck\ candidates with low-quality images on DSS red
plates\footnote{\url{http://stdatu.stsci.edu/dss/}} or without SDSS
information, or with low signal-to-noise ratio in RASS, were imaged to
the depth needed for finding an optical counterpart and for
determination of a photometric redshift.  Candidates with galaxy
concentrations in SDSS or with high signal-to-noise ratio in RASS were
sent for spectroscopic confirmation.

\section{Optical identifications and redshift measurements}
\label{sec:opt}

\subsection{Surface number density of galaxies}
\label{sec:galdens}

\begin{figure*}
  \centering
   \includegraphics[width=0.3\linewidth]{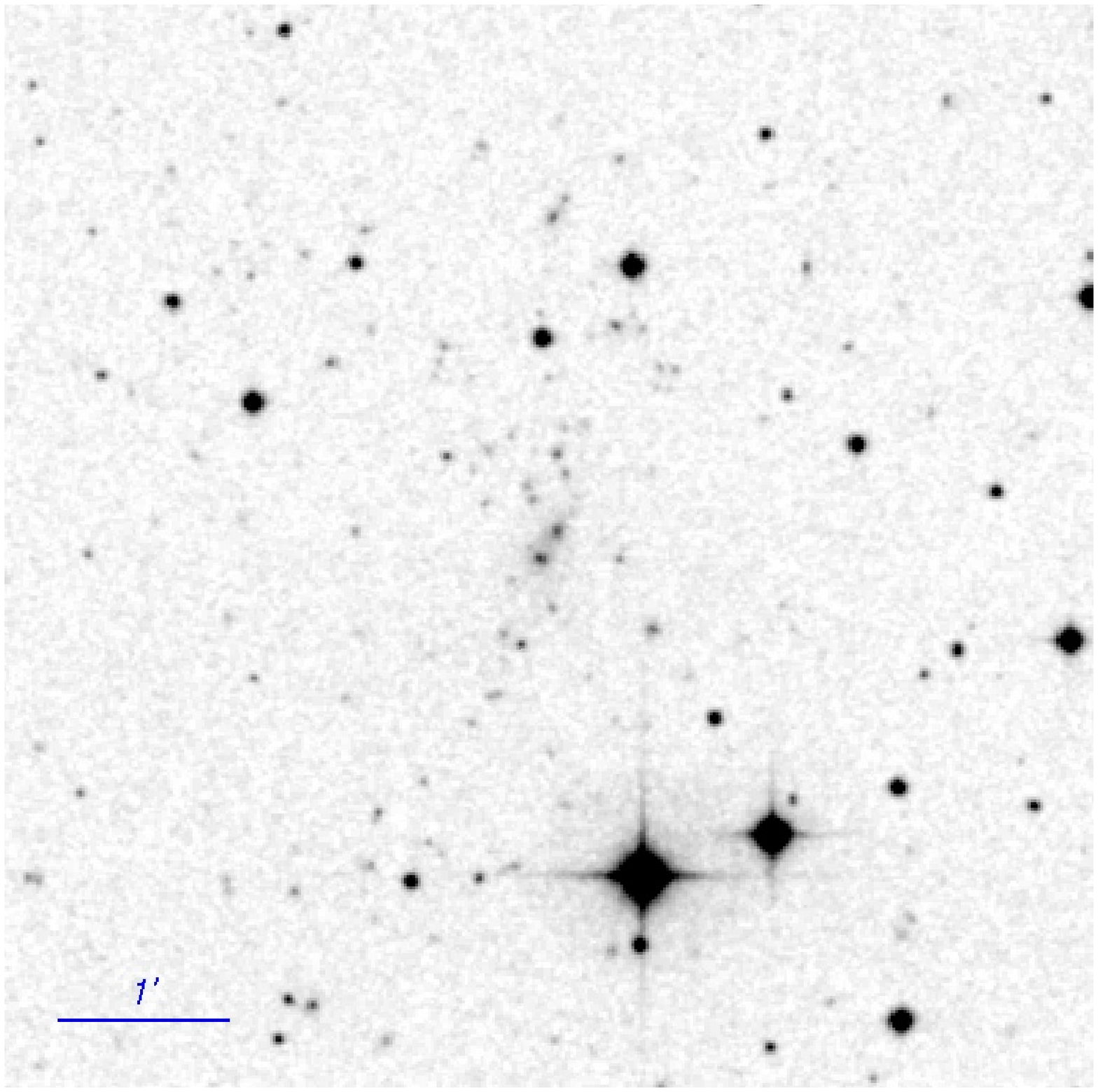}
   ~~~
   \includegraphics[width=0.3\linewidth]{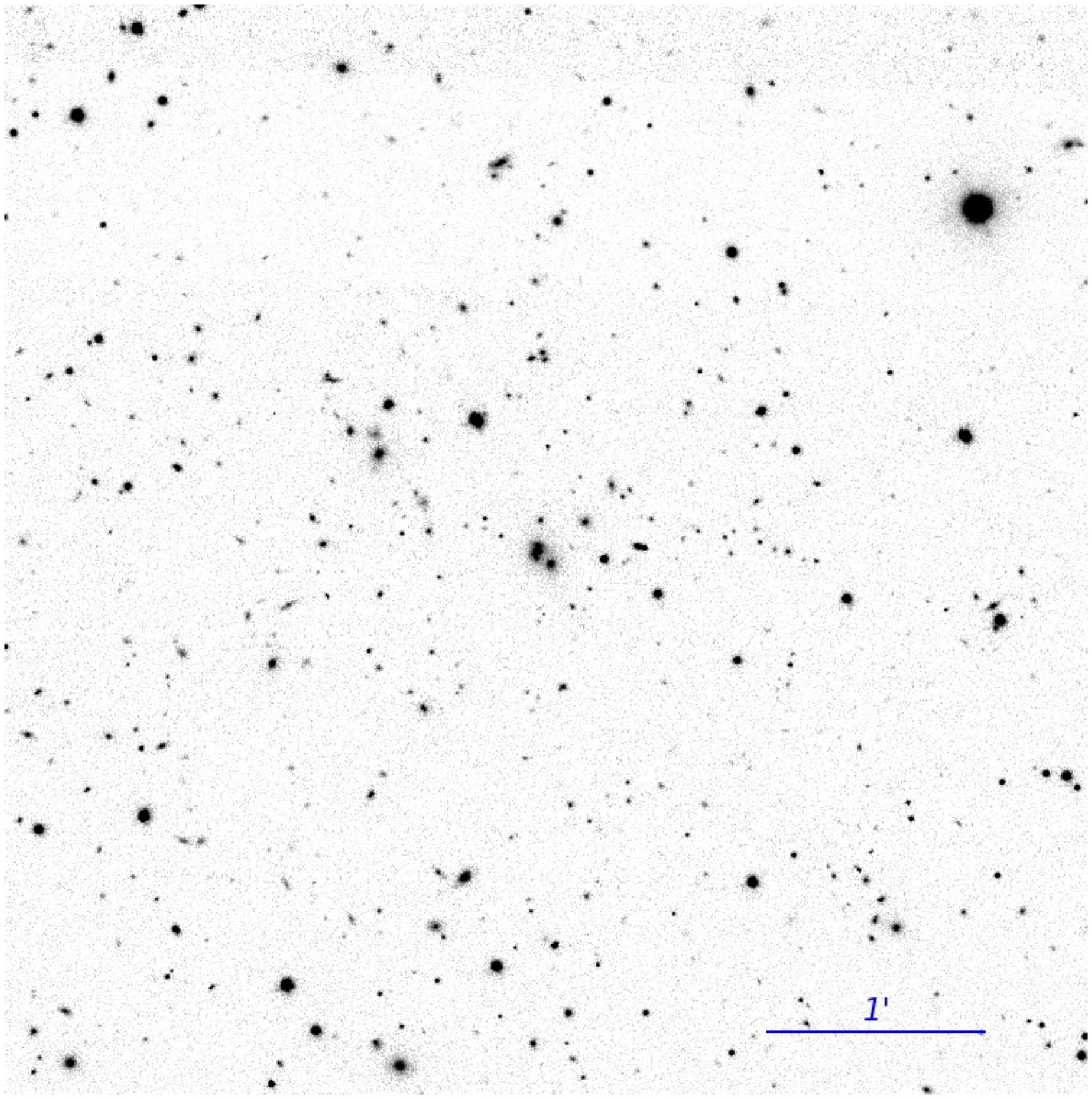}
   ~~~
   \includegraphics[width=0.3\linewidth]{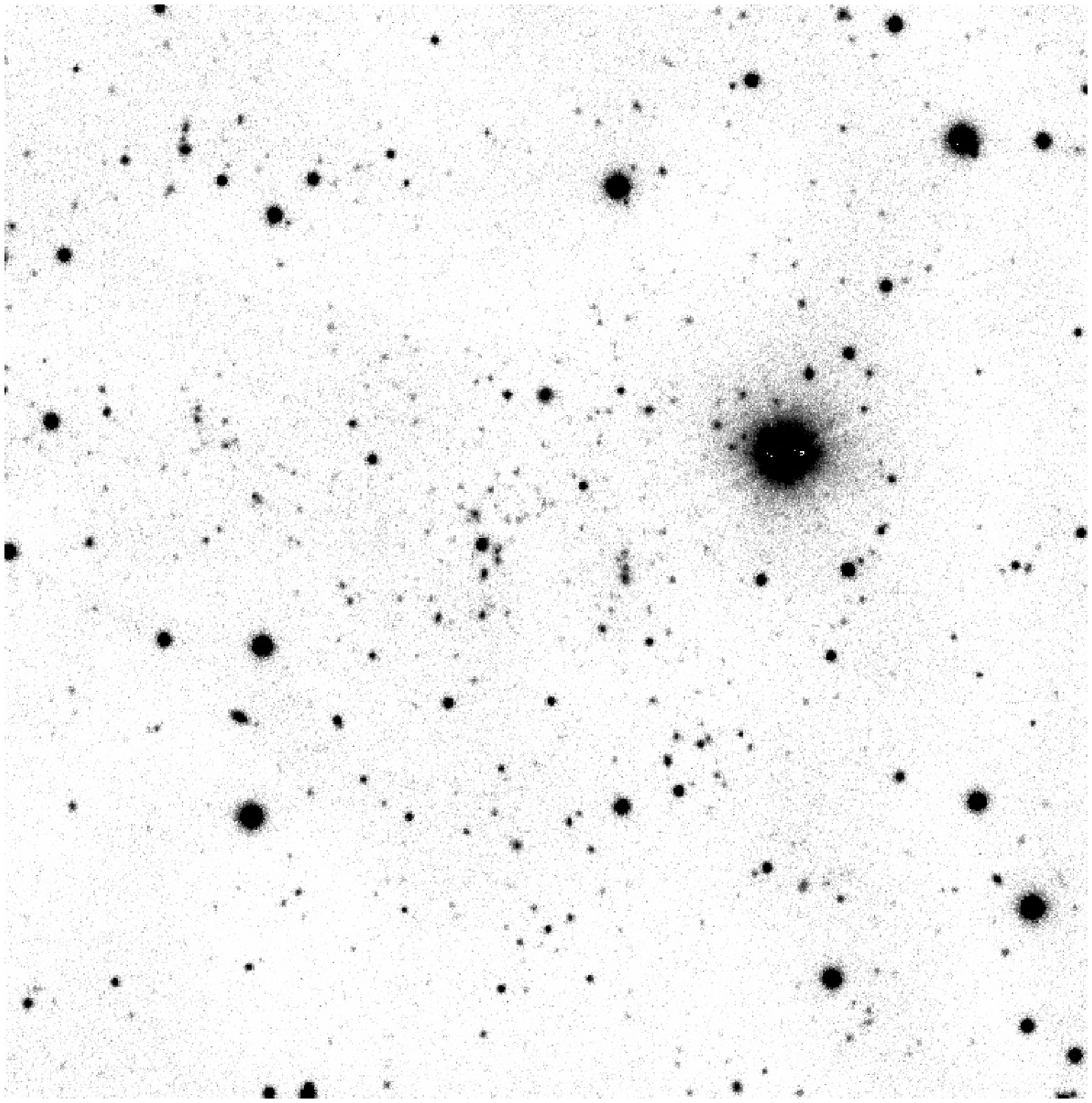}
   \medskip
   
   \caption{Examples of \Planck\ galaxy clusters. \,\, {\it Left\/}:
     DSS red image of PSZ1 G101.52-29.96 at $z=0.227$. \,\, {\it
       Centre\/}: SDSS $i^\prime$-band image of PSZ1 G054.94-33.37 at
     $z=0.392$. \,\, {\it Right\/}: deeper RTT150 $i^\prime$-band
     image of PSZ1 G108.26+48.66 at $z=0.674$. The images are
     approximately 1.5~Mpc on a side at the redshift of the cluster.}
   \label{fig:optimg}
\end{figure*}

The most straightforward way to identify galaxy clusters in the
optical is through detection of an enhanced surface number density of
galaxies.  The fields of all cluster candidates were inspected on DSS
red plates and then in SDSS images if available.  On DSS plates,
clusters can be identified at redshifts up to about $0.3$, while SDSS
images allow reliable identifications at redshifts up to redshift
around $0.6$. To identify more distant clusters, deeper direct images
are necessary.  For clusters at $z\approx0.8$, therefor a 1.5-m class
telescope, images in $r^\prime$ and $i^\prime$ with approximately one
hour exposures are needed.  Clusters at $z\approx1$ and higher should
be observed at larger telescopes.  Examples are given in
Fig.~\ref{fig:optimg}.  At lower Galactic latitudes, where the density
of stars is higher, the galaxy surface density enhancement associated
with clusters is easier to detect when stars are excluded\footnote{In
  our work star--galaxy separation in optical images was performed
  using \texttt{SExtractor} \citep{1996A&AS..117..393B}}.


\begin{figure*}
  \centering
  
  \includegraphics[width=0.4\linewidth]{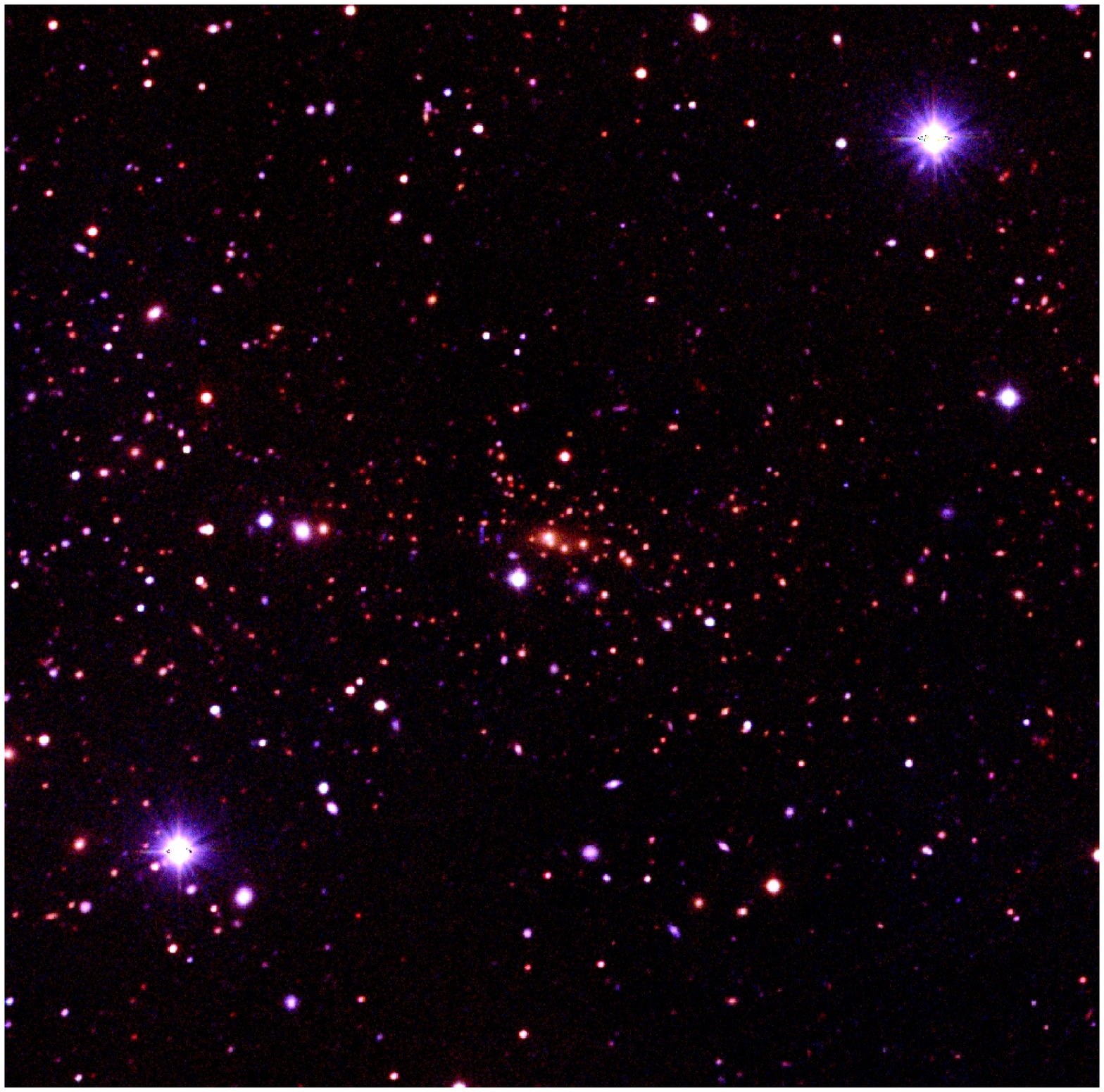}
  ~
  \includegraphics[width=0.4\linewidth]{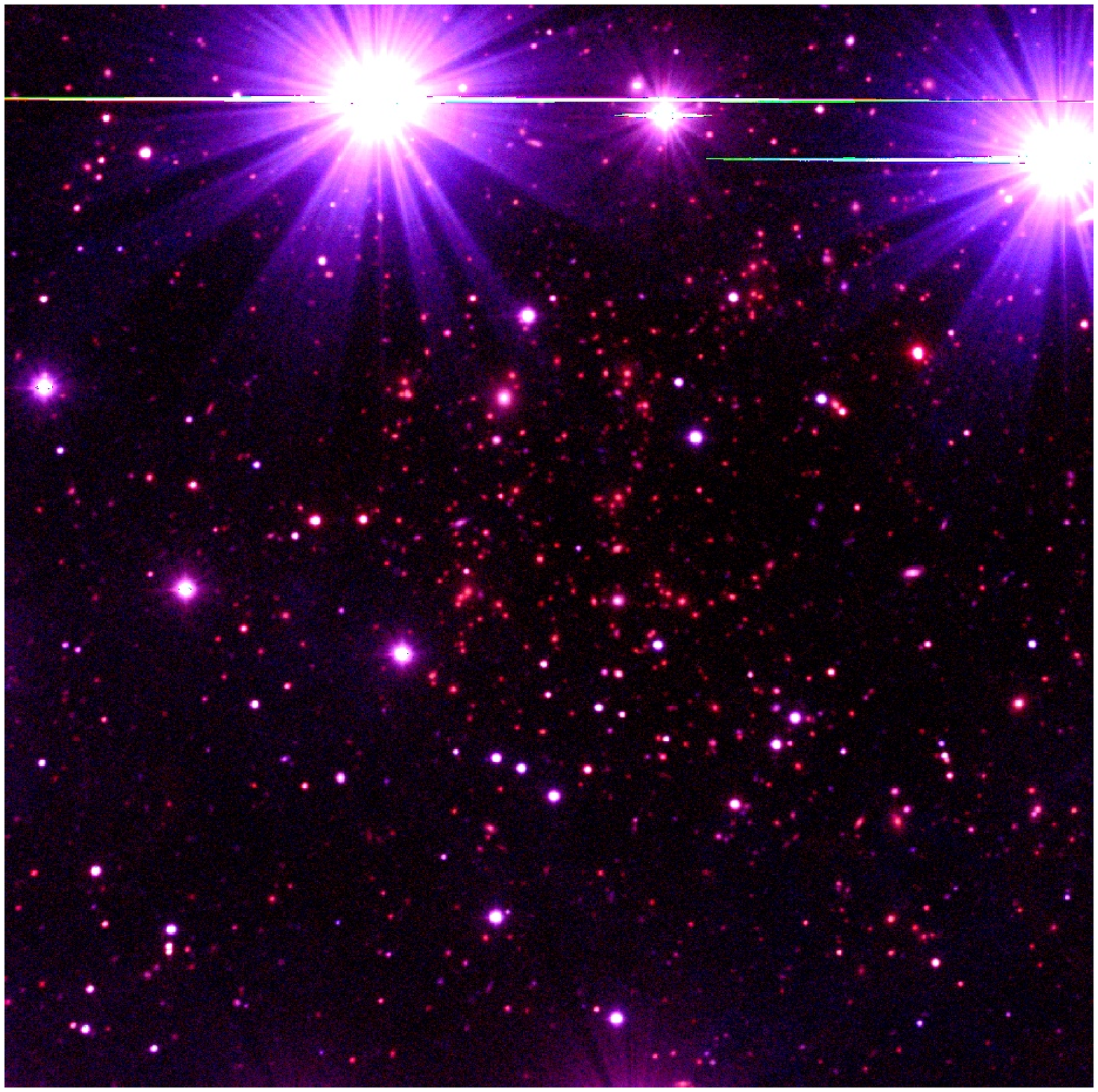}

  \vskip 2mm
  
  ~\includegraphics[width=0.4\linewidth]{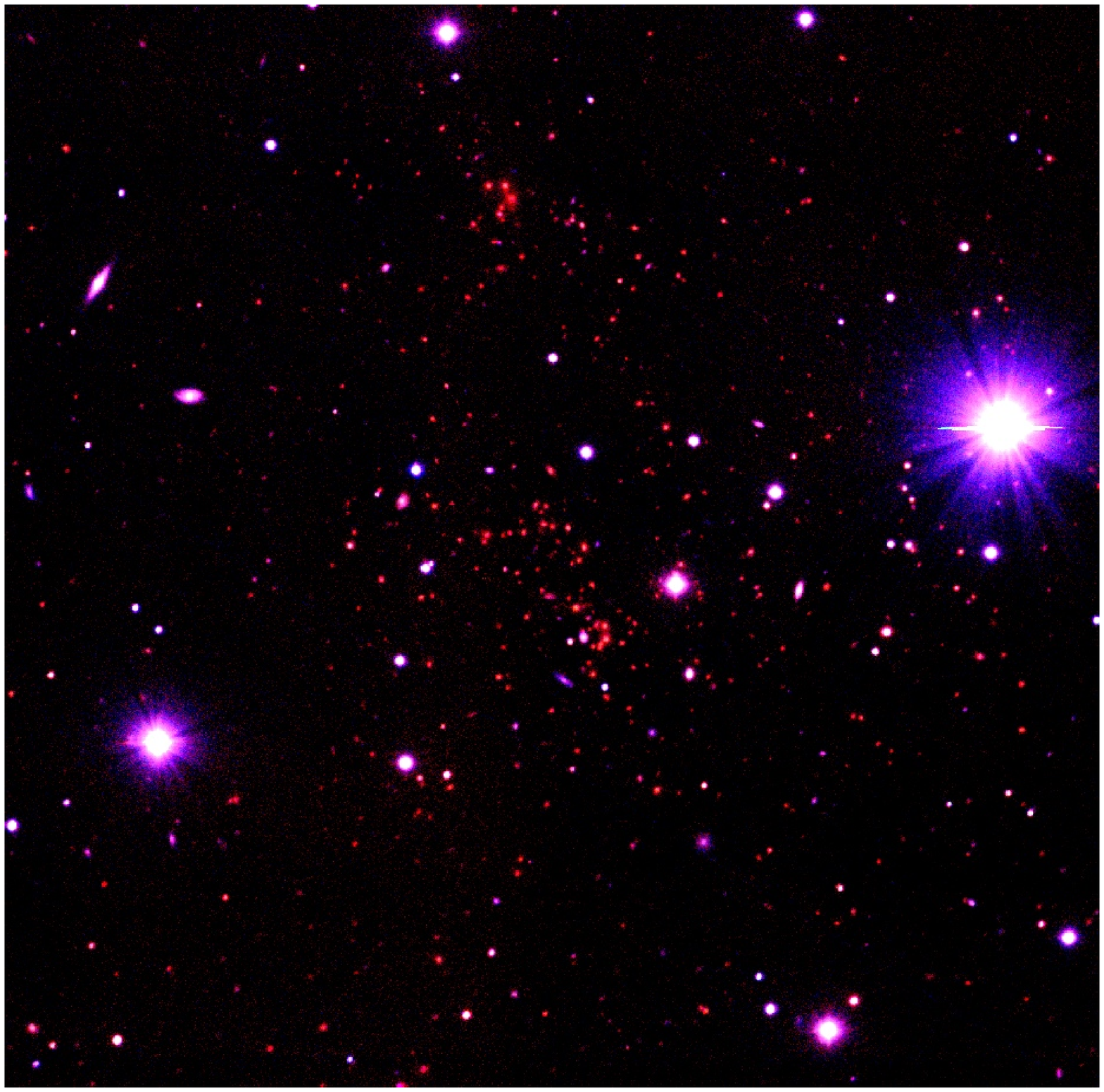}
  ~
  \includegraphics[width=0.4\linewidth]{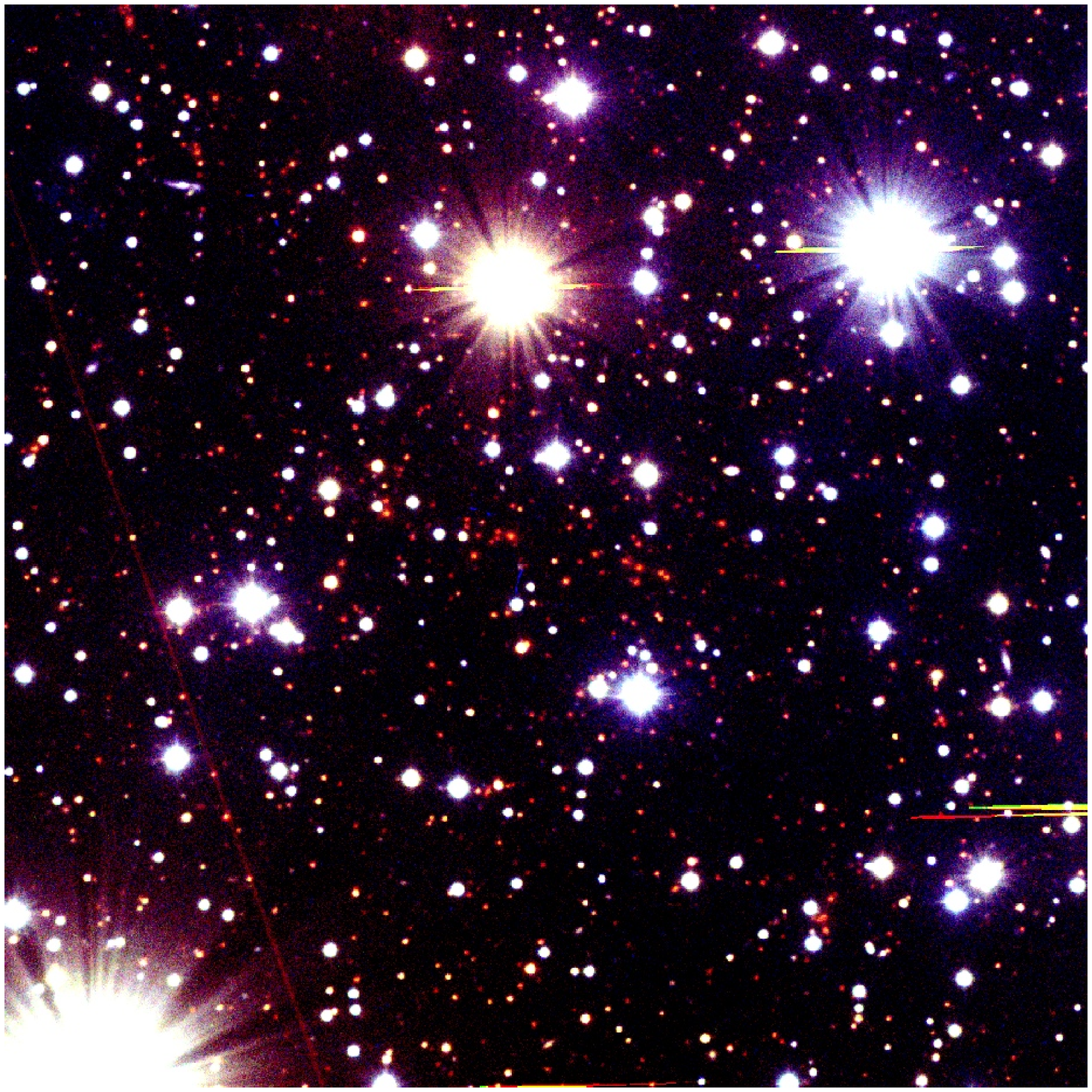}
  \medskip
  
  \caption{Pseudo colour ($g^\prime r^\prime i^\prime$) RTT150 images
    of \Planck\ clusters, with the colour map adjusted to emphasize
    the red sequence of galaxies in the centres of clusters. \,\, {\it
      Upper left\/}: PSZ1 G098.24-41.15, $z=0.436$.  \,\, {\it Upper
      right\/}: PSZ1 G100.18-29.68, $z=0.485$. \,\, {\it Lower
      left\/}: PSZ1 G138.11+42.03, $z=0.496$.  \,\, {\it Lower
      right\/}: PSZ1 G209.80+10.23, $z=0.677$.}
  \label{fig:gri}
\end{figure*}

\subsection{Red sequence}
\label{sec:rs}

Spectroscopic redshifts are obtained most effectively from spectra of
the brightest galaxies in the central part of the cluster. In order to
correctly identify the brightest cluster members, foreground galaxies
must be securely rejected. That can be done through detection of a red
sequence in the galaxy colour-magnitude relation, formed by early-type
cluster member galaxies \citep[][]{2000AJ....120.2148G}.

\begin{figure}
  \centering
   \includegraphics[width=1.0\linewidth,bb=23 163 550 682]{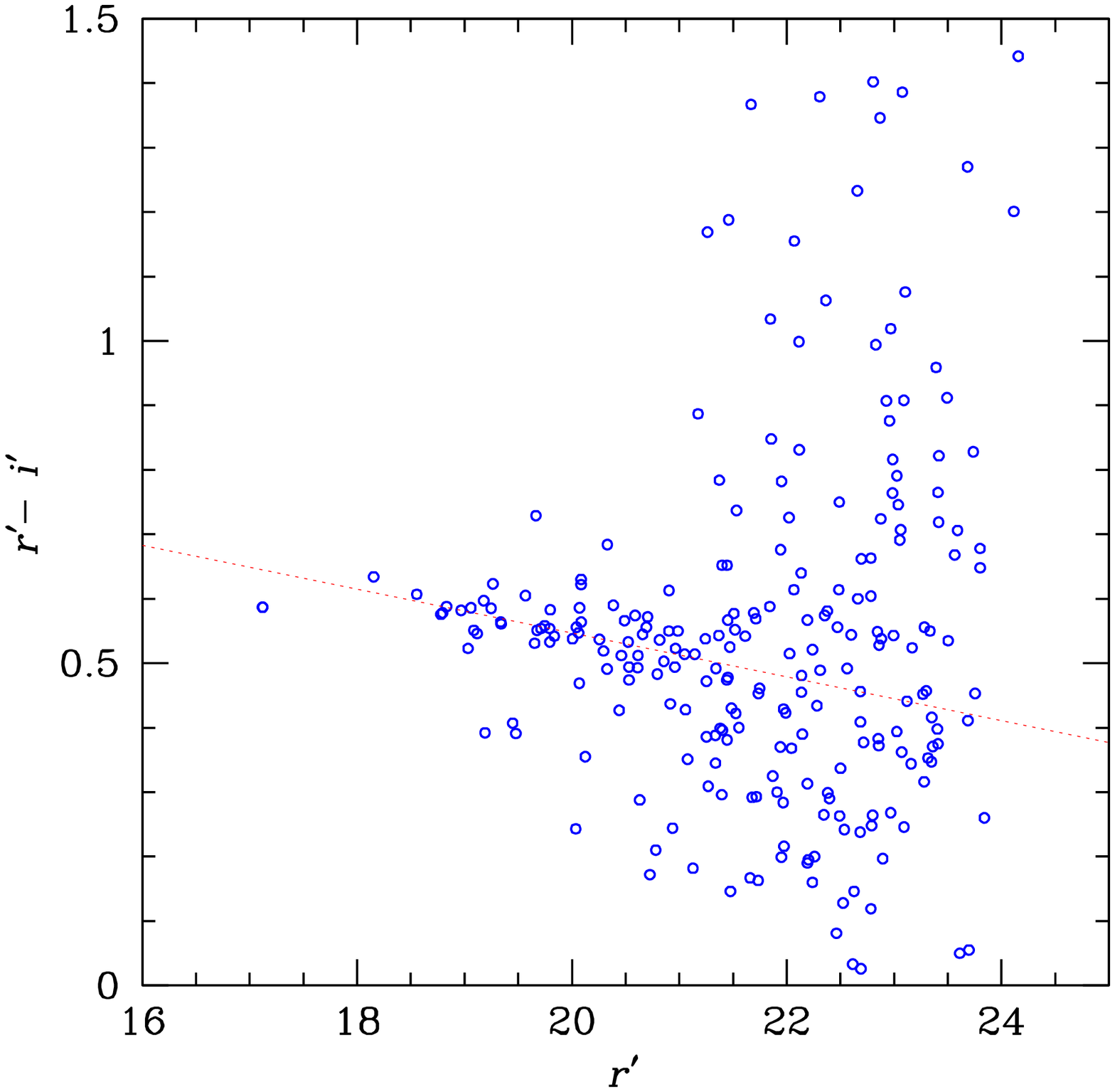}
   \caption{Colour-magnitude diagram of galaxies near the centre of
     cluster PSZ1 G101.52-29.96 at $z=0.227$ (see the left panel in
     Fig.~\ref{fig:optimg}). The best fit red sequence is shown with
     a dotted line.}
   \label{fig:p819rs}
\end{figure}

Red galaxies, which form the cluster red sequence, are clearly visible
near the centre of clusters, as shown in Fig.~\ref{fig:gri}, where
pseudo colour $g^\prime r^\prime i^\prime$ images of a few {\it Planck}
clusters obtained at RTT150 are presented.  Figure~\ref{fig:p819rs}
gives an example of a colour-magnitude diagram of galaxies near the
centre of the field of a cluster at $z=0.227$ (PSZ1 G101.52-29.96,
left panel in Fig.~\ref{fig:optimg}). One can see that observations of
red-sequence galaxies in clusters provide an efficient way to identify
clusters and their member galaxies.

\begin{figure}
  \centering
   \includegraphics[width=1.0\linewidth,bb=23 163 550 682]{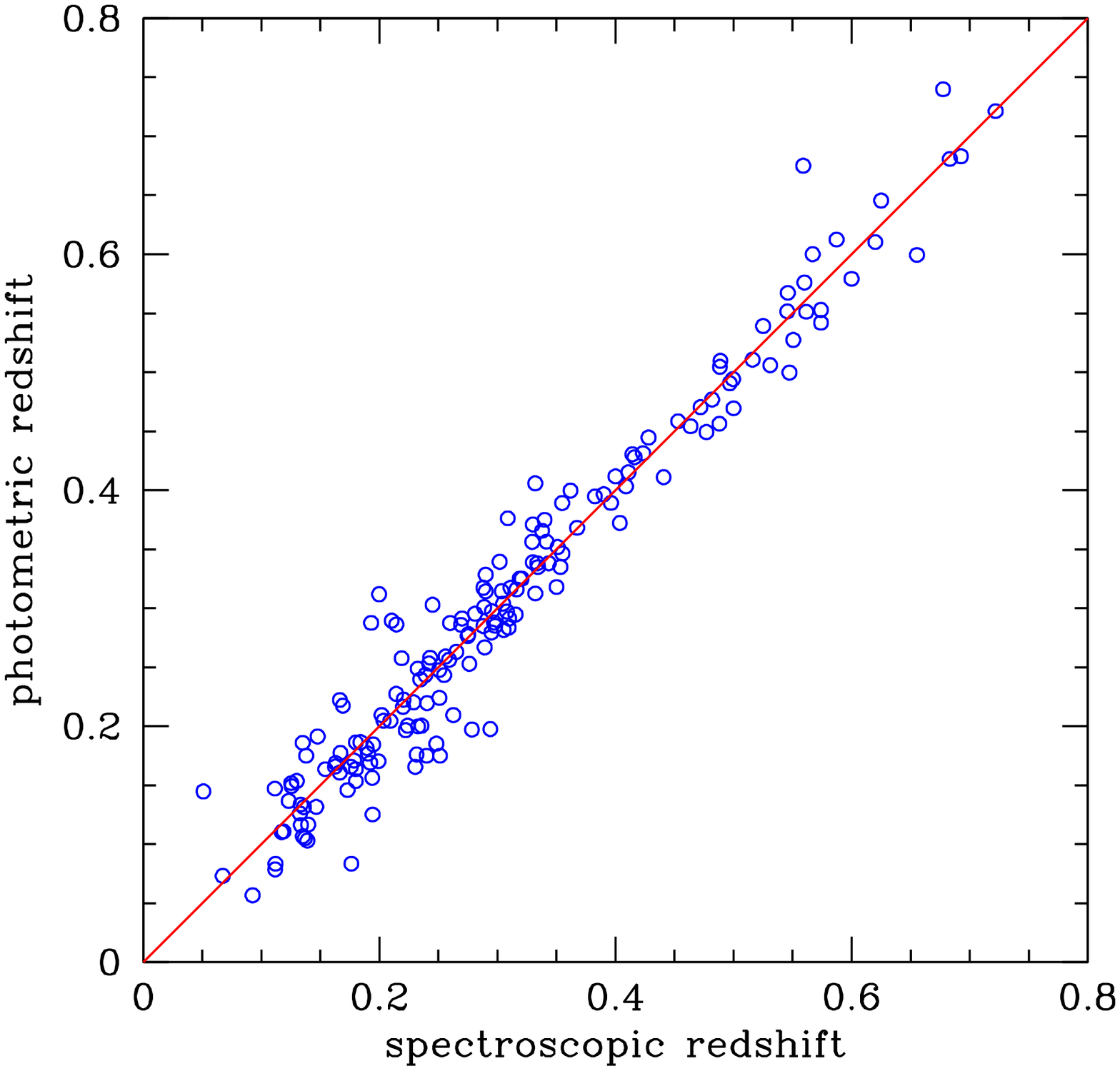}
   \caption{Comparison of photometric redshifts based on red-sequence
     colours with spectroscopic redshifts. The points show the
       data for clusters from the 400d cluster survey \citep{400d}, which
       were used for photometric redshift calibration.}
   \label{fig:zphot}
\end{figure}

Observed red-sequence colours can be used to estimate photometric
redshifts. For the purposes of this work, photometric redshifts were
calibrated using the data of optical observations obtained earlier
during the construction of the 400\,deg$^2$ {\it ROSAT} PSPC galaxy
cluster survey \citep{400d}. The results of this calibration are shown
in Fig.~\ref{fig:zphot}. The accuracy of the redshift estimates is
$\delta z/(1+z)=0.027$.
These preliminary estimates were used mainly for planning
observations. Since the accuracy of photometric redshifts is
insufficient for accurate cluster mass function measurements, the
redshifts for all confirmed clusters should be measured
spectroscopically.

Cluster members identified from a red sequence typically form a
well-defined concentration of galaxies, with one brightest cD galaxy
in the centre (see, e.g., Fig.~\ref{fig:optimg} and the upper left
panel in Fig.~\ref{fig:gri}). In some cases, multiple cD-like galaxies
are observed in the centre of a concentration (see, e.g., the
right-hand panels in Fig.~\ref{fig:gri}). In some cases, the
concentration of cluster galaxies is observed to have two or more
peaks, with a few cD-like galaxies at the centre of each peak (see,
e.g., the lower left panel in Fig.~\ref{fig:gri}). In our work, the
optical centres of galaxy clusters were determined from the positions
of cD galaxies found in the centres of these concentrations.

Some galaxy clusters are not associated with prominent enhancements in
galaxy surface density. Instead, clusters may be dominated in the
optical by one giant central galaxy.  These rare objects, called
fossil groups
\citep{1994Natur.369..462P,1999ApJ...520L...1V,2003MNRAS.343..627J,
  2010ApJ...708.1376V}, are much less massive than the typical cluster
detected by \Planck.  For example, all 12~fossil systems identified in
the emph{400d} survey \citep{2010ApJ...708.1376V} have X-ray
luminosities below $10^{44}$~erg~s$^{-1}$, with corresponding masses
below $3\times10^{14}$\,M$_{\mathord\odot}$ \citep[see,
e.g.,][]{av09a}.  Nevertheless, objects similar to fossil groups are
detected in the \Planck\ survey at low redshifts.  These objects and
their member galaxies can still be reliably identified by their red
sequences. An example is discussed in Sect.~~\ref{sec:ex}.

There are also \Planck\ SZ sources where two or more clusters at
different redshifts are projected on the sky within a few arcminutes. 
In those cases, it is not easy to determine the
contribution of each cluster to the SZ signal detected by \Planck. These,
and some other special cases, are discussed in detail below
(Sect.~\ref{sec:notes},~\ref{sec:complicated}).

\subsection{Spectroscopic redshift measurements}
\label{sec:spec}

Once cluster members are identified through a red sequence, the
redshift of the cluster as a whole can be determined from the
brightest galaxies near the centre of the cluster.  For regular
clusters, we measured the redshift of the dominant cD galaxy. For less
regular clusters, we measured redshifts of 3--5 brightest galaxies,
selected using the cluster red sequence observations.

\begin{figure*}
  \includegraphics[width=180mm]{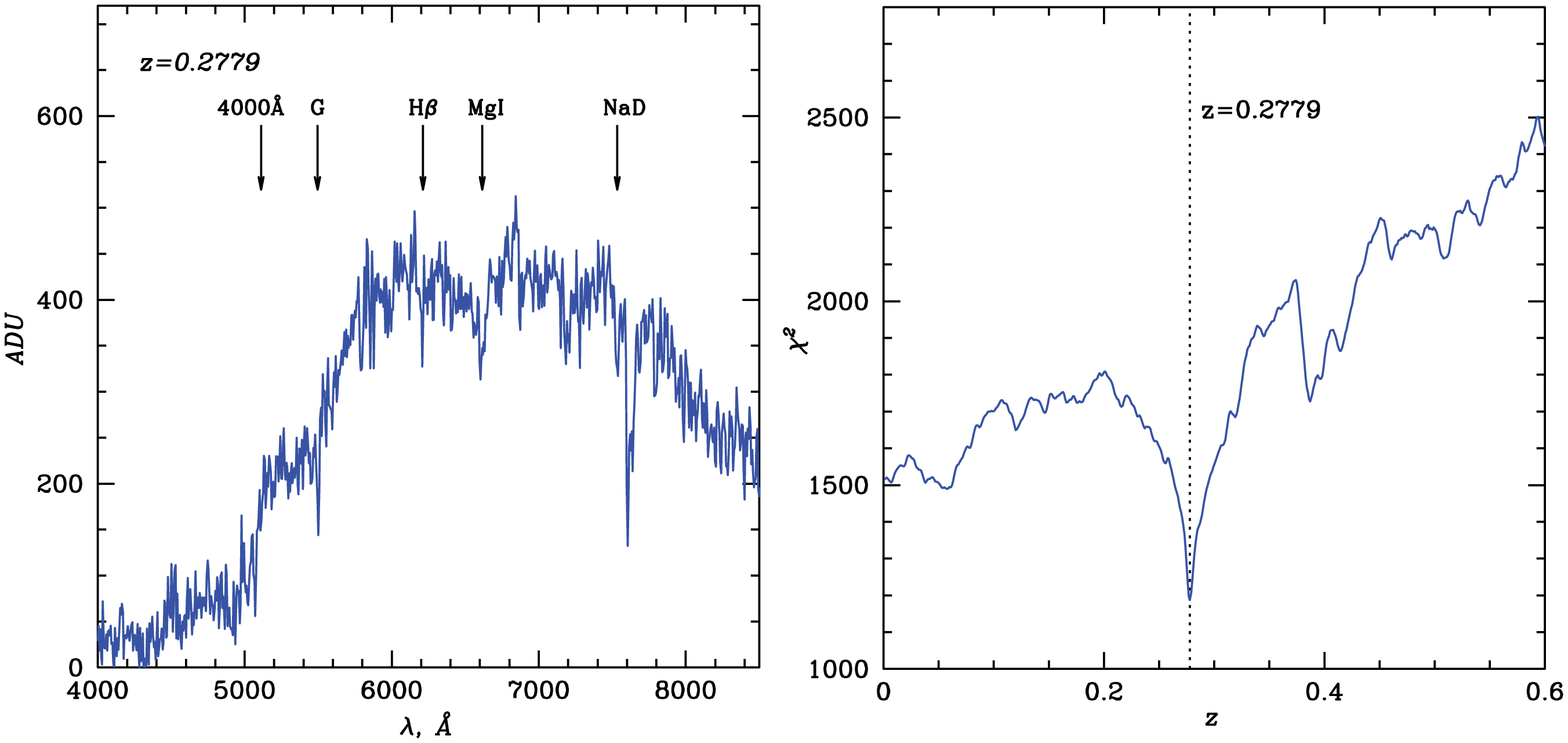}
  \caption{Example of the spectrum of the central elliptical galaxy of
    cluster at $z=0.278$ (PSZ1 G066.24+20.82), obtained at RTT150
    using the TFOSC spectrometer (left), together with $\chi^2$ from
    the cross-correlation with an elliptical galaxy template spectrum
    (right).}
  \label{fig:zspec}
\end{figure*}

High signal-to-noise ratio is not necessary for accurate spectroscopic
redshift determination. Even if individual spectral lines are not
well-identified, the redshift can be determined accurately by
cross-correlation with an elliptical galaxy template spectrum.
Figure~\ref{fig:zspec}, for example, shows the spectrum of the
brightest central galaxy in a cluster at $z=0.278$ obtained with the
TFOSC spectrometer, along with $\chi^2$ as a function of $z$ from the
cross-correlation with an elliptical galaxy template spectrum.

\section{Observations}
\label{sec:obs}

Deep multi-colour observations were obtained for all cluster
candidates except those unambiguously detected in SDSS. Images were
obtained with the RTT150 telescope and the TFOSC instrument through
Sloan $g^\prime r^\prime i^\prime$ filters, typically with 1800\,s
exposures per filter. Longer exposures were used for cluster
candidates with brightest galaxies fainter than $m_{r^\prime}\approx
21$, i.e., which may be located at redshifts $z>0.7$
\citep[e.g.,][]{160d}.
Images were obtained in a series of 300 or 600\,s exposures with
$\approx10\arcsec$--$30\arcsec$ pointing offsets between exposures.
Standard CCD calibrations were applied using
\texttt{Iraf}\footnote{\url{http://iraf.noao.edu/}}
software. Individual images in each filter were then aligned and
combined. With these data, galaxy clusters can be efficiently
identified at redshifts up to $z\approx1$.

We identified cluster members from a red sequence. Clusters whose
photometry indicated $z<0.4$ were observed spectroscopically with the
RTT150. In some cases, clusters whose photometry indicated redshifts
above $0.4$ were observed spectroscopically with the BTA 6-m telescope
and the SCORPIO spectrometer (Sect.~\ref{sec:rtt150}).

\section{Results}
\label{sec:res} 

\begin{figure}
  \centering
  \includegraphics[width=88mm]{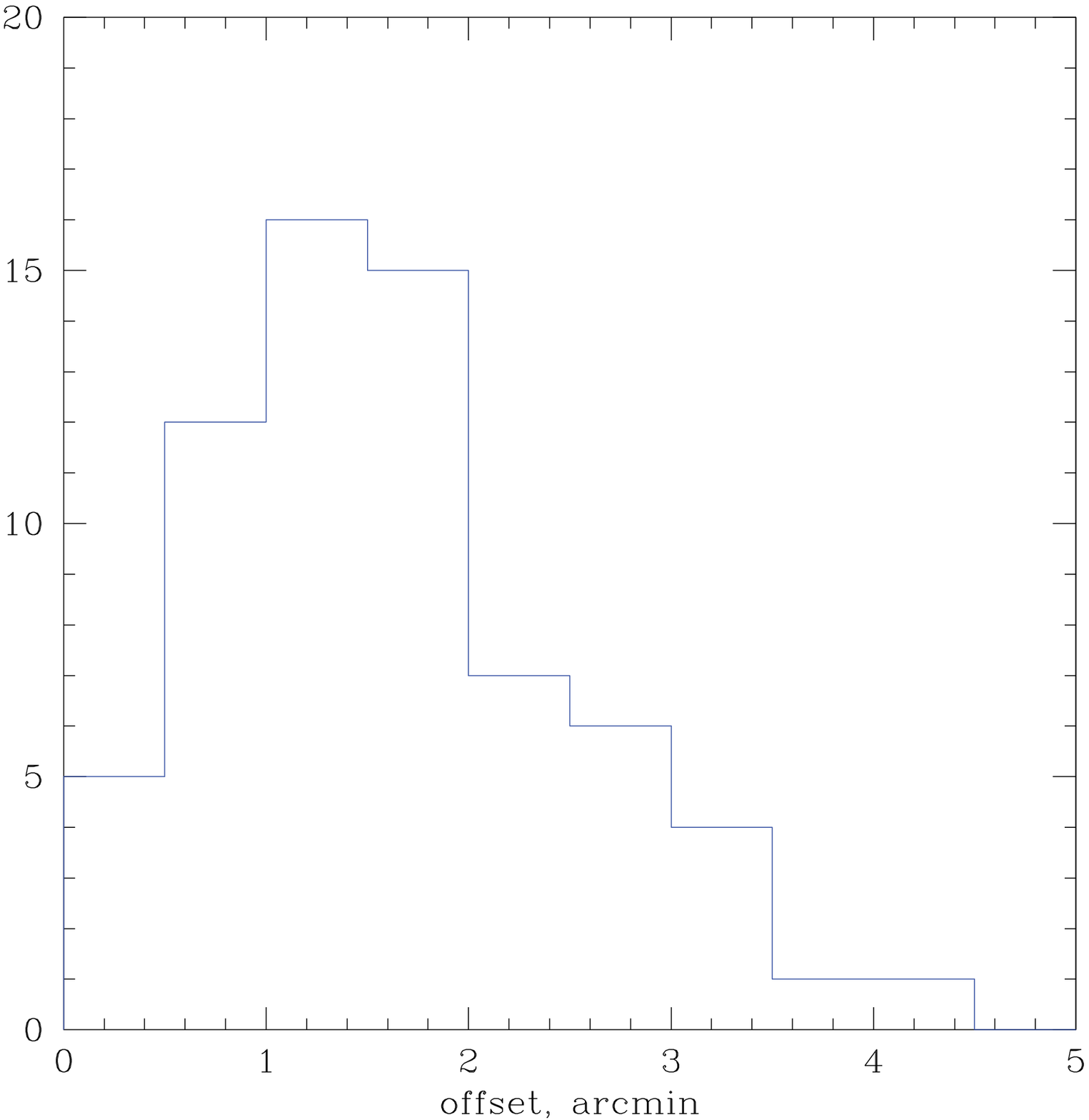}
    \caption{The distribution of cluster optical centres offsets
    relative to their SZ position measured with {\it Planck}.}
  \label{fig:off}
\end{figure}

The list of \Planck\ clusters from the PSZ1 catalogue observed with
the RTT150 is given in Table~\ref{tab:clcat}. Clusters identified with
the RTT150 that were below the PSZ1 signal-to-noise detection limit of
4.5$\,\sigma$, and thus not included in the PSZ1 catalogue, are given
in Table~\ref{tab:clnotincat}. Coordinates are given of the cluster
optical centres, calculated from the position of cD galaxies. The
distribution of cluster optical centre offsets from the SZ positions
measured with \Planck\ is shown in Fig.~\ref{fig:off}, and is
consistent with a two-dimensional Gaussian distribution of width
$2\arcmin$, in agreement with the \Planck\ positional accuracy given
in \citep{PSZcat13}.

In total, deep direct images of more than one hundred fields in
multiple filters were obtained. Forty-seven clusters newly identified
using the RTT150 imaging data are marked in Tables~\ref{tab:clcat}
(41) and \ref{tab:clnotincat} (6). Redshifts of 65
\Planck\ clusters were measures spectroscopically, including 12
at high redshift measured with the 6-m BTA telescope. Two of these
redshifts were published in \cite{2013A&A...550A.128P}, and 26 were
included in the PSZ1 catalogue \citep{PSZcat13}.

Table~\ref{tab:clcat} also gives new spectroscopic redshifts measured
at the RTT150 for clusters in the PSZ1 catalogue with previously
published photometric redshifts (e.g., \citealt{2012ApJS..199...34W}),
and new photometric redshifts estimated from the RTT150 data of
14~clusters still lacking spectroscopic redshifts.

Below, we give examples of the \Planck\ SZ cluster identifications
with RTT150 data, showing cases where a 1.5-m-class telescope can be
used to identify clusters, while DSS and SDSS data are insufficient for
these purposes.  We then give notes on some individual objects from
Table~\ref{tab:clcat}, discuss complicated cases where more data in SZ
or X-rays are needed, and identify some probable false clusters among
the objects observed in our programme.

\begin{table*}
\begingroup
\caption{Clusters from the PSZ1 catalogue observed with the RTT150.} 
\label{tab:clcat}
\nointerlineskip
\vskip -3mm
\footnotesize
\setbox\tablebox=\vbox{
   \newdimen\digitwidth
   \setbox0=\hbox{\rm 0}
   \digitwidth=\wd0
   \catcode`*=\active
   \def*{\kern\digitwidth}
   \newdimen\signwidth
   \setbox0=\hbox{+}
   \signwidth=\wd0
   \catcode`!=\active
   \def!{\kern\signwidth}
\halign{\hbox to 1.5in{#\leaderfil}\tabskip=2em&
   \hfil#\hfil\tabskip=1em&
   \hfil#\hfil\tabskip=2em&
   \hfil#\hfil&
   \hfil#\hfil&
   #\hfil\tabskip=0pt\cr
\noalign{\doubleline}
\omit&\multispan2\hfil Position (J2000)\hfil\cr
\noalign{\vskip -3pt}
\omit&\multispan2\hrulefill\cr
\omit\hfil Name\hfil&R. A.&Decl.&$z$&New ID&\omit\hfil Notes\hfil\cr
\noalign{\vskip 3pt\hrule\vskip 5pt}
PSZ1 G034.78$+$31.71& 16 58 38.6& $+$15 19 13& 0.480*&&\cr 
PSZ1 G037.69$-$46.90& 21 50 36.8& $-$16 22 29& 0.263*& +&\cr 
PSZ1 G038.71$+$24.47& 17 31 59.1& $+$15 40 42& 0.0836&&\cr 
PSZ1 G042.33$+$17.46& 18 04 16.1& $+$16 02 16& 0.50*\rlap{$^{\rm a}$}*& +&\cr 
PSZ1 G045.54$+$16.26& 18 14 13.3& $+$18 17 04& 0.24*\rlap{$^{\rm a}$}*& +&\cr 
PSZ1 G046.13$+$30.75& 17 17 05.7& $+$24 04 18& 0.569\rlap{$^{\rm c,d}$}*& +&\cr 
PSZ1 G048.22$-$65.03& 23 09 51.0& $-$18 19 57& 0.42*\rlap{$^{\rm a}$}*& +&\cr 
PSZ1 G049.04$+$25.26& 17 43 30.4& $+$24 44 19& 0.141\rlap{$^{\rm c}$}*&&  ACO 2279\cr 
PSZ1 G050.07$-$27.29& 20 58 53.0& $+$01 24 11& 0.334\rlap{$^{\rm c}$}*&&\cr 
PSZ1 G054.94$-$33.37& 21 28 23.4& $+$01 35 37& 0.392*&&\cr 
PSZ1 G055.72$+$17.58& 18 25 30.0& $+$27 44 27& 0.194*&&\cr 
PSZ1 G056.76$-$11.60& 20 18 48.2& $+$15 07 25& 0.122\rlap{$^{\rm b}$}*&& ZwCl 2016.6+1457\cr 
PSZ1 G059.20$+$32.92& 17 20 49.9& $+$35 20 50& 0.383\rlap{$^{\rm c}$}*& +&\cr 
PSZ1 G059.98$-$18.66& 20 50 27.1& $+$13 45 18& 0.221*&&\cr 
PSZ1 G060.12$+$11.42& 18 58 46.0& $+$29 15 34& 0.30*\rlap{$^{\rm a}$}*& +&\cr 
PSZ1 G063.92$-$16.75& 20 52 51.7& $+$17 54 23& 0.393\rlap{$^{\rm b}$}*& +&\cr 
PSZ1 G065.13$+$57.53& 15 16 02.0& $+$39 44 26& 0.685\rlap{$^{\rm b,d}$}*& +&\cr 
PSZ1 G066.01$-$23.30& 21 19 26.2& $+$15 21 06& 0.248\rlap{$^{\rm b}$}*& +&\cr 
PSZ1 G066.24$+$20.82& 18 27 26.5& $+$38 14 50& 0.278*&&\cr 
PSZ1 G066.41$+$27.03& 17 56 52.6& $+$40 08 07& 0.576\rlap{$^{\rm c}$}*&&\cr 
PSZ1 G067.02$-$20.80& 21 14 01.4& $+$17 43 14& 0.334*& +&\cr 
PSZ1 G069.92$-$18.89& 21 15 09.6& $+$21 01 10& 0.308*& +&\cr 
PSZ1 G070.91$+$49.26& 15 56 25.6& $+$44 40 42& 0.607\rlap{$^{\rm b,d}$}*&& \cr 
PSZ1 G071.57$-$37.96& 22 17 15.8& $+$09 03 10& 0.25*\rlap{$^{\rm a}$}*&& ACO 2429\cr 
PSZ1 G073.22$+$67.57& 14 20 40.3& $+$39 55 11& 0.609\rlap{$^{\rm c,d}$}*& +&\cr 
PSZ1 G076.44$+$23.53& 18 28 21.8& $+$48 04 29& 0.169*& +&\cr 
PSZ1 G079.33$+$28.33& 18 02 09.6& $+$51 37 11& 0.204*&& ZwCl 1801.2+5136\cr 
PSZ1 G079.88$+$14.97& 19 23 12.1& $+$48 16 14& 0.0998\rlap{$^{\rm b}$}& +&\cr 
PSZ1 G080.11$-$77.29& 00 15 24.4& $-$17 30 34& 0.43*\rlap{$^{\rm a}$}*& +&\cr 
PSZ1 G084.04$+$58.75& 14 49 00.2& $+$48 33 28& 0.731\rlap{$^{\rm b,d}$}*& +&\cr 
PSZ1 G084.41$-$12.43& 21 37 46.6& $+$35 35 51& 0.273\rlap{$^{\rm b}$}*& +&\cr 
PSZ1 G085.71$+$10.67& 20 03 13.4& $+$51 20 51& 0.0805\rlap{$^{\rm b}$}& +&\cr 
PSZ1 G087.47$+$37.65& 16 57 20.7& $+$58 28 54& 0.113*&&\cr 
PSZ1 G090.82$+$44.13& 16 03 35.1& $+$59 11 41& 0.269\rlap{$^{\rm c}$}*&&\cr 
PSZ1 G092.27$-$55.73& 23 44 23.1& $+$03 04 42& 0.349\rlap{$^{\rm c}$}*&& \cr 
PSZ1 G095.37$+$14.42& 20 14 29.2& $+$61 23 30& 0.119*& +&\cr 
PSZ1 G098.24$-$41.15& 23 34 24.1& $+$17 59 23& 0.436*&&\cr 
PSZ1 G100.18$-$29.68& 23 21 02.9& $+$29 12 52& 0.485*&& \cr 
PSZ1 G101.52$-$29.96& 23 26 26.6& $+$29 21 44& 0.227*&&\cr 
PSZ1 G102.00$+$30.69& 17 47 13.0& $+$71 23 17& 0.214\rlap{$^{\rm b}$}*&& ZwCl 1748.0+7125\cr 
PSZ1 G106.15$+$25.76& 18 56 51.7& $+$74 55 53& 0.588\rlap{$^{\rm b,d}$}*& +&\cr 
PSZ1 G107.66$-$58.31& 00 19 37.6& $+$03 37 49& 0.267\rlap{$^{\rm c}$}*&&\cr 
PSZ1 G108.18$-$11.53& 23 22 29.7& $+$48 46 30& 0.336\rlap{$^{\rm b,d}$}*& & \cr
PSZ1 G108.26$+$48.66& 14 27 04.6& $+$65 39 47& 0.674\rlap{$^{\rm b,d}$}*& +&\cr 
PSZ1 G109.14$-$28.02& 23 53 12.7& $+$33 16 11& 0.457\rlap{$^{\rm c,d}$}*& +& \cr 
PSZ1 G114.81$-$11.80& 00 01 14.7& $+$50 16 33& 0.228\rlap{$^{\rm b}$}*& +&\cr 
PSZ1 G115.70$+$17.51& 22 26 28.3& $+$78 16 58& 0.50*\rlap{$^{\rm a}$}*&&\cr 
PSZ1 G118.40$+$42.23& 13 42 02.2& $+$74 25 21& 0.478*& +&\cr 
PSZ1 G121.09$+$57.02& 12 59 33.0& $+$60 04 12& 0.344*&&\cr 
PSZ1 G123.55$-$10.34& 00 55 24.4& $+$52 29 20& 0.107\rlap{$^{\rm b}$}*& +&\cr 
PSZ1 G127.02$+$26.21& 05 58 02.3& $+$86 13 50& 0.574\rlap{$^{\rm b,d}$}*& +&\cr 
PSZ1 G129.07$-$24.12& 01 20 00.0& $+$38 25 18& 0.425\rlap{$^{\rm b}$}*& +&\cr 
PSZ1 G130.26$-$26.53& 01 23 39.6& $+$35 53 58& 0.216*&& ZwCl 0120.8+3538\cr 
PSZ1 G134.31$-$06.57& 02 10 25.1& $+$54 34 09& 0.44*\rlap{$^{\rm a}$}*& +&\cr 
PSZ1 G134.64$-$11.77& 02 02 37.8& $+$49 26 28& 0.207\rlap{$^{\rm b}$}*&&\cr 
PSZ1 G138.11$+$42.03& 10 28 09.7& $+$70 34 26& 0.496*& +&\cr 
PSZ1 G139.61$+$24.20& 06 21 48.9& $+$74 42 06& 0.267*&&\cr 
PSZ1 G141.73$+$14.22& 04 41 05.8& $+$68 13 16& 0.833\rlap{$^{\rm b,d}$}*& +&\cr 
PSZ1 G149.38$-$36.86& 02 21 33.8& $+$21 21 58& 0.170*&& ACO 344\cr 
\noalign{\vskip 3pt\hrule\vskip 3pt}}}
\endPlancktablewide
\tablenote {{\rm a}} Estimated from RTT photometric data.\par
\tablenote {{\rm b}} Not published in the PSZ1 \citep{PSZcat13}\par
\tablenote {{\rm c}} Spectroscopic redshift. The redshift given in the PSZ1 catalogue \citep{PSZcat13} is photometric.\par
\tablenote {{\rm d}} Measured at the BTA 6-m telescope of SAO RAS.\par
\endgroup
\end{table*}

\begin{table*}
\begingroup
\leftline{{\bf Table 1 --- continued}}
\nointerlineskip
\footnotesize
\setbox\tablebox=\vbox{
   \newdimen\digitwidth
   \setbox0=\hbox{\rm 0}
   \digitwidth=\wd0
   \catcode`*=\active
   \def*{\kern\digitwidth}
   \newdimen\signwidth
   \setbox0=\hbox{+}
   \signwidth=\wd0
   \catcode`!=\active
   \def!{\kern\signwidth}
\halign{\hbox to 1.5in{#\leaderfil}\tabskip=2em&
   \hfil#\hfil\tabskip=1em&
   \hfil#\hfil\tabskip=2em&
   \hfil#\hfil&
   \hfil#\hfil&
   #\hfil\tabskip=0pt\cr
\noalign{\doubleline}
\omit&\multispan2\hfil Position (J2000)\hfil\cr
\noalign{\vskip -3pt}
\omit&\multispan2\hrulefill\cr
\omit\hfil Name\hfil&R. A.&Decl.&$z$&New ID&\omit\hfil Notes\hfil\cr
\noalign{\vskip 3pt\hrule\vskip 5pt}
PSZ1 G151.80$-$48.06& 02 08 06.5& $+$10 27 19& 0.202\rlap{$^{\rm b}$}*&& ACO 307\cr 
PSZ1 G155.95$-$72.13& 01 30 42.1& $-$11 51 39& 0.620\rlap{$^{\rm e}$}*& +&\cr 
PSZ1 G156.88$+$13.48& 05 45 41.3& $+$55 30 40& 0.235\rlap{$^{\rm b}$}*&& \cr
PSZ1 G157.44$+$30.34& 07 48 54.3& $+$59 42 06& 0.45*\rlap{$^{\rm a}$}*&& [ATZ98] B100\cr 
PSZ1 G157.84$+$21.23& 06 40 32.7& $+$57 45 36& 0.43*\rlap{$^{\rm a}$}*& +&\cr 
PSZ1 G170.22$+$09.74& 06 03 16.8& $+$42 14 42& 0.228\rlap{$^{\rm b}$}*&& 1RXS J060313.4+42123\cr 
PSZ1 G171.01$+$15.93& 06 35 47.9& $+$44 10 15& 0.281\rlap{$^{\rm b}$}*& +&\cr 
PSZ1 G172.93$+$21.31& 07 07 38.1& $+$44 19 55& 0.331*& +&\cr 
PSZ1 G183.26$+$12.25& 06 43 09.9& $+$31 50 55& 0.85*\rlap{$^{\rm a}$}*& +&\cr 
PSZ1 G188.56$-$68.86& 02 11 44.2& $-$17 00 46& 0.174\rlap{$^{\rm b}$}*&& ACO 2985\cr 
PSZ1 G205.56$-$55.75& 03 15 22.0& $-$18 12 22& 0.31*\rlap{$^{\rm a}$}*& +&\cr 
PSZ1 G209.80$+$10.23& 07 22 23.8& $+$07 24 31& 0.677*& +&\cr 
PSZ1 G216.77$+$09.25& 07 31 20.3& $+$00 49 24& 0.273\rlap{$^{\rm c}$}*& +&\cr 
PSZ1 G222.75$+$12.81& 07 54 58.6& $-$02 41 29& 0.369*& +&\cr 
PSZ1 G223.04$-$20.27& 05 54 37.3& $-$17 44 35& 0.19*\rlap{$^{\rm a}$}*&& ACO 551\cr 
PSZ1 G224.01$-$11.14& 06 30 55.3& $-$14 51 00& 0.62*\rlap{$^{\rm a}$}*&&\cr 
PSZ1 G338.97$+$35.62& 14 52 42.2& $-$18 35 05& 0.297\rlap{$^{\rm b}$}*& +&\cr 
\noalign{\vskip 3pt\hrule\vskip 3pt}}}
\endPlancktablewide
\tablenote {{\rm a}} Estimated from RTT photometric data.\par
\tablenote {{\rm b}} Not published in the PSZ1 \citep{PSZcat13}\par
\tablenote {{\rm c}} Spectroscopic redshift.  The redshift given in the PSZ1 catalogue \citep{PSZcat13} is photometric.\par
\tablenote {{\rm d}} Measured at the BTA 6-m telescope of SAO RAS.\par
\tablenote {{\rm e}} Redshift from the PSZ1 catalogue, given here for completeness.\par
\endgroup
\end{table*}

\begin{table*}
\begingroup
\caption{Clusters below the $4.5\,\sigma$ limit of the PSZ1 catalogue
  observed with the RTT150.}
 \label{tab:clnotincat} 
\nointerlineskip
\footnotesize
\setbox\tablebox=\vbox{
   \newdimen\digitwidth
   \setbox0=\hbox{\rm 0}
   \digitwidth=\wd0
   \catcode`*=\active
   \def*{\kern\digitwidth}
   \newdimen\signwidthf
   \setbox0=\hbox{+}
   \signwidth=\wd0
   \catcode`!=\active
   \def!{\kern\signwidth}
\halign{\hbox to 1.55in{#\leaderfil}\tabskip=2em&
   \hfil#\hfil\tabskip=1em&
   \hfil#\hfil\tabskip=2em&
   \hfil#\hfil&
   \hfil#\hfil\tabskip=0pt\cr
\noalign{\doubleline}
\omit&\multispan2\hfil Position (J2000)\hfil\cr
\noalign{\vskip -3pt}
\omit&\multispan2\hrulefill\cr
\omit\hfil Name\hfil&R. A.&Decl.&$z$&New ID\cr
\noalign{\vskip 3pt\hrule\vskip 5pt}
PLCK G201.42$-$56.60&03 08 18.1& $-$16 26 01& \dots&+\cr
PLCK G210.76$+$08.02&  07 16 08.1& $+$05 34 01& 0.296&+\cr
PLCK G201.03$+$29.90&  08 23 55.5& $+$22 43 51& 0.668\rlap{$^{\rm a}$}&+\cr
PLCK G244.13$+$26.11&  09 24 06.2& $-$12 19 05& \dots&+\cr
PLCK G164.28$+$52.61&  10 16 19.9& $+$50 10 29& 0.488\cr
PLCK G041.62$+$57.43&  15 18 22.3& $+$27 10 18& \dots&+\cr
PLCK G050.55$-$25.00&20 51 56.4& $+$02 56 43& \dots&+\cr
\noalign{\vskip 3pt\hrule\vskip 3pt}}}
\endPlancktablewide
\tablenote {{\rm a}} Measured at the BTA 6-m telescope of SAO RAS.\par
\endgroup
\end{table*}

\begin{figure*}
  \centering
  \includegraphics[height=0.4\linewidth]{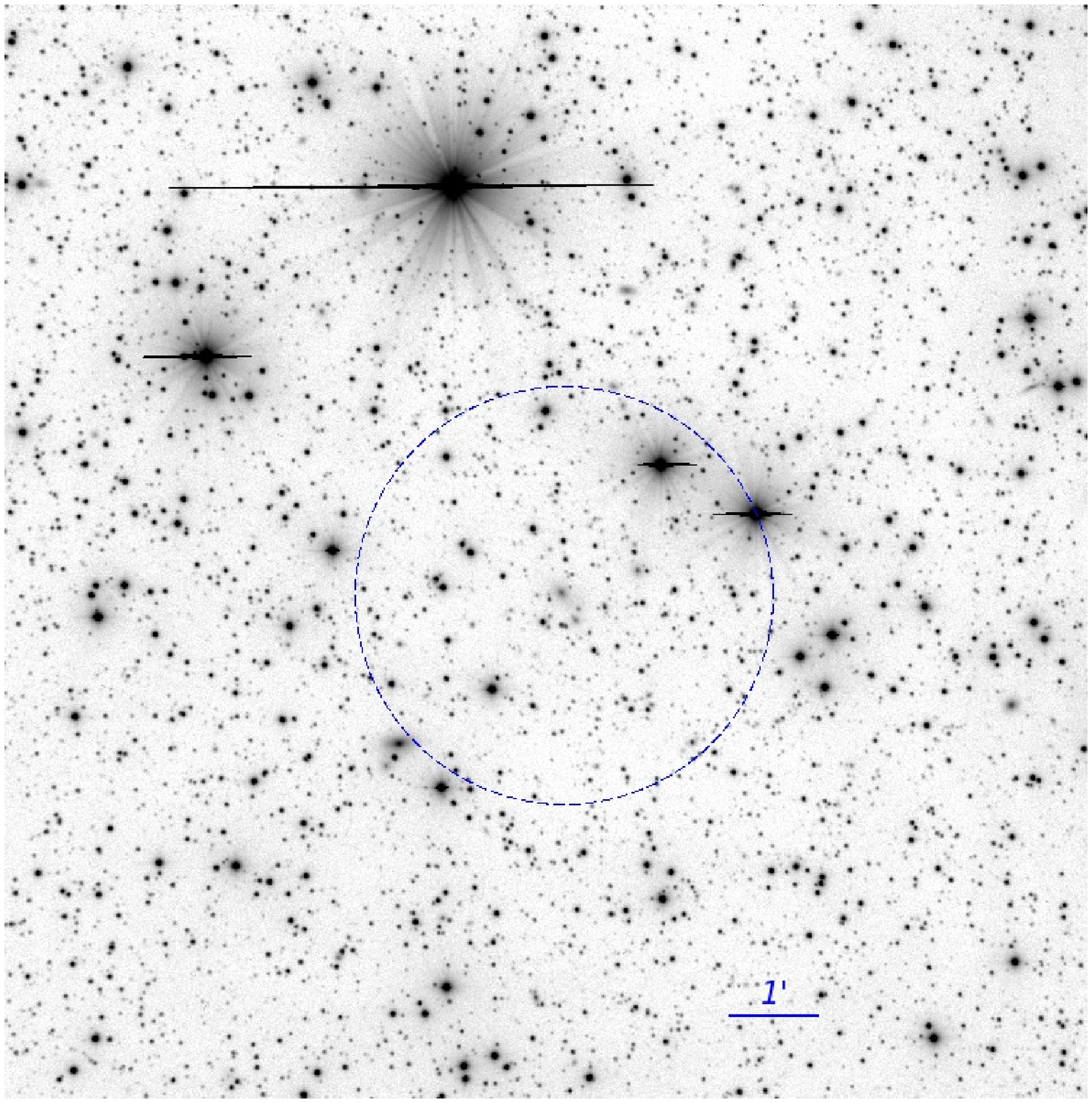}
  ~~~
  \includegraphics[height=0.4\linewidth]{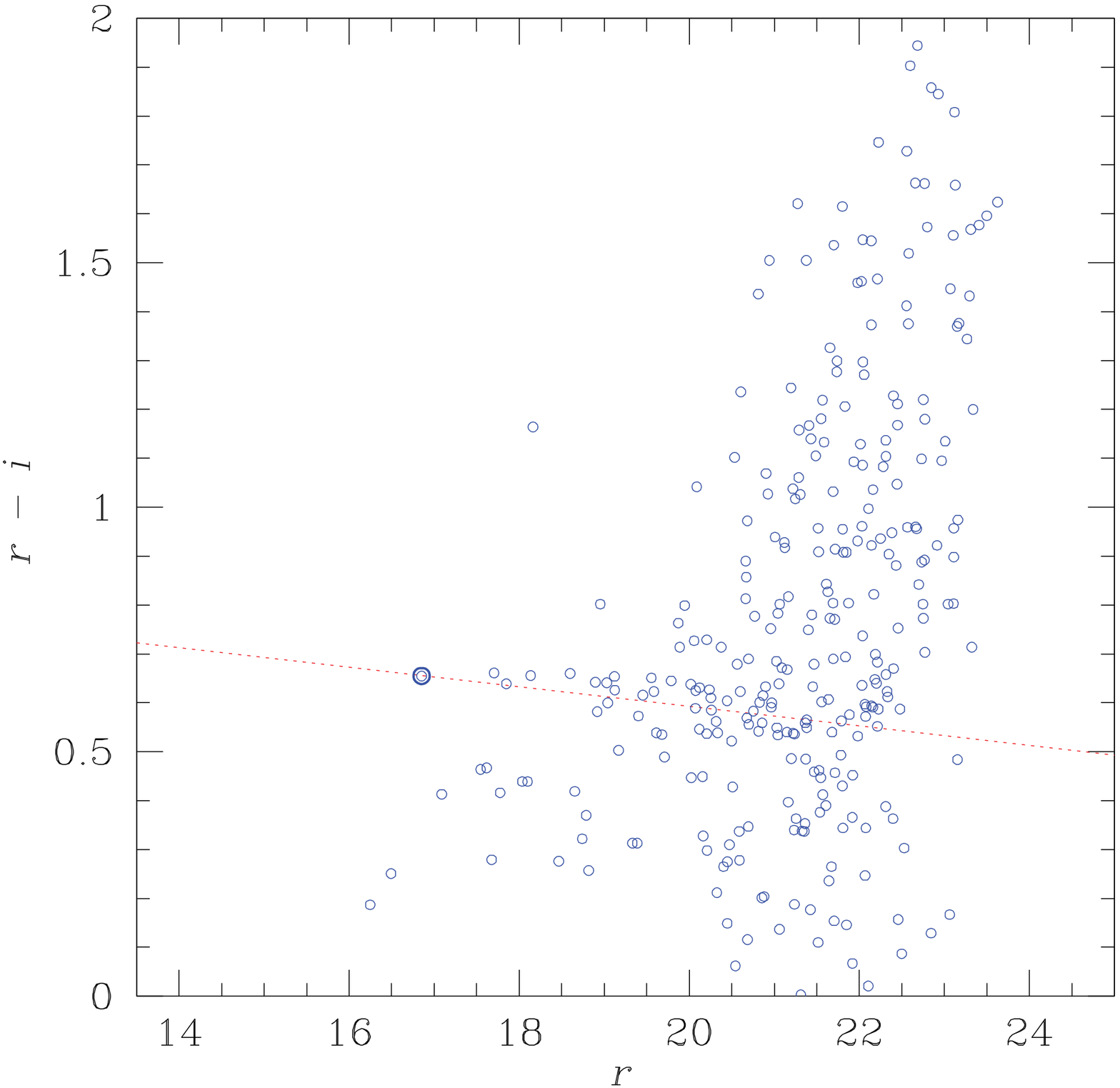}
  \medskip
  
  \caption{PSZ1 G060.12$+$11.42, a cluster at low Galactic latitude
    ($b\approx-11\pdeg4$). {\it Left\/}: $i^\prime$-band RTT150
    image. {\it Right\/}: colour-magnitude diagram of extended objects
    near the cluster centre in the area enclosed by the dashed circle
    in the left panel. The dotted line indicates the red sequence.}
  \label{fig:g06012}
  \medskip
  ~
\end{figure*}


\begin{figure*}
  \centering
  \includegraphics[height=0.4\linewidth]{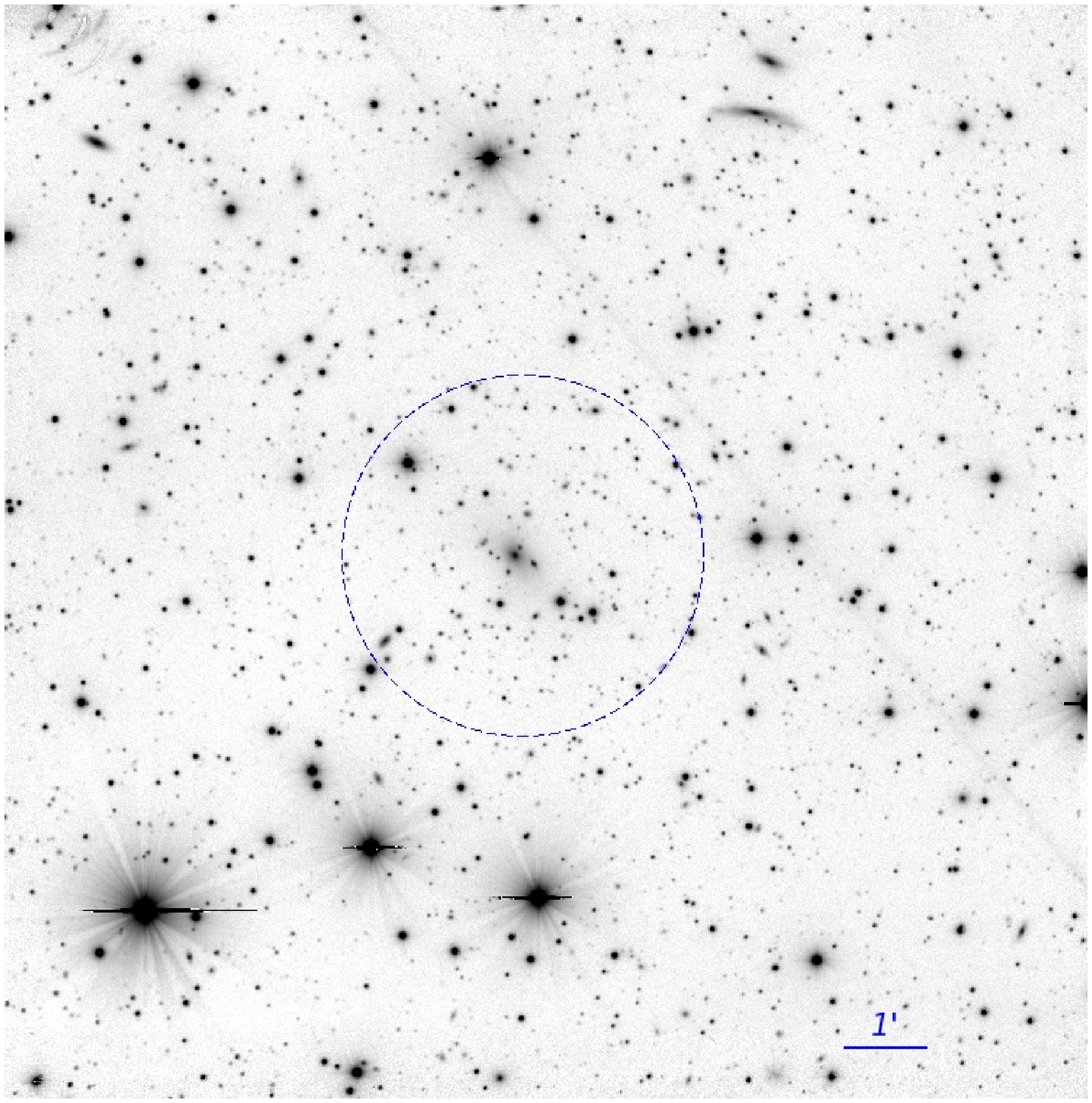}
  ~~~
  \includegraphics[height=0.4\linewidth]{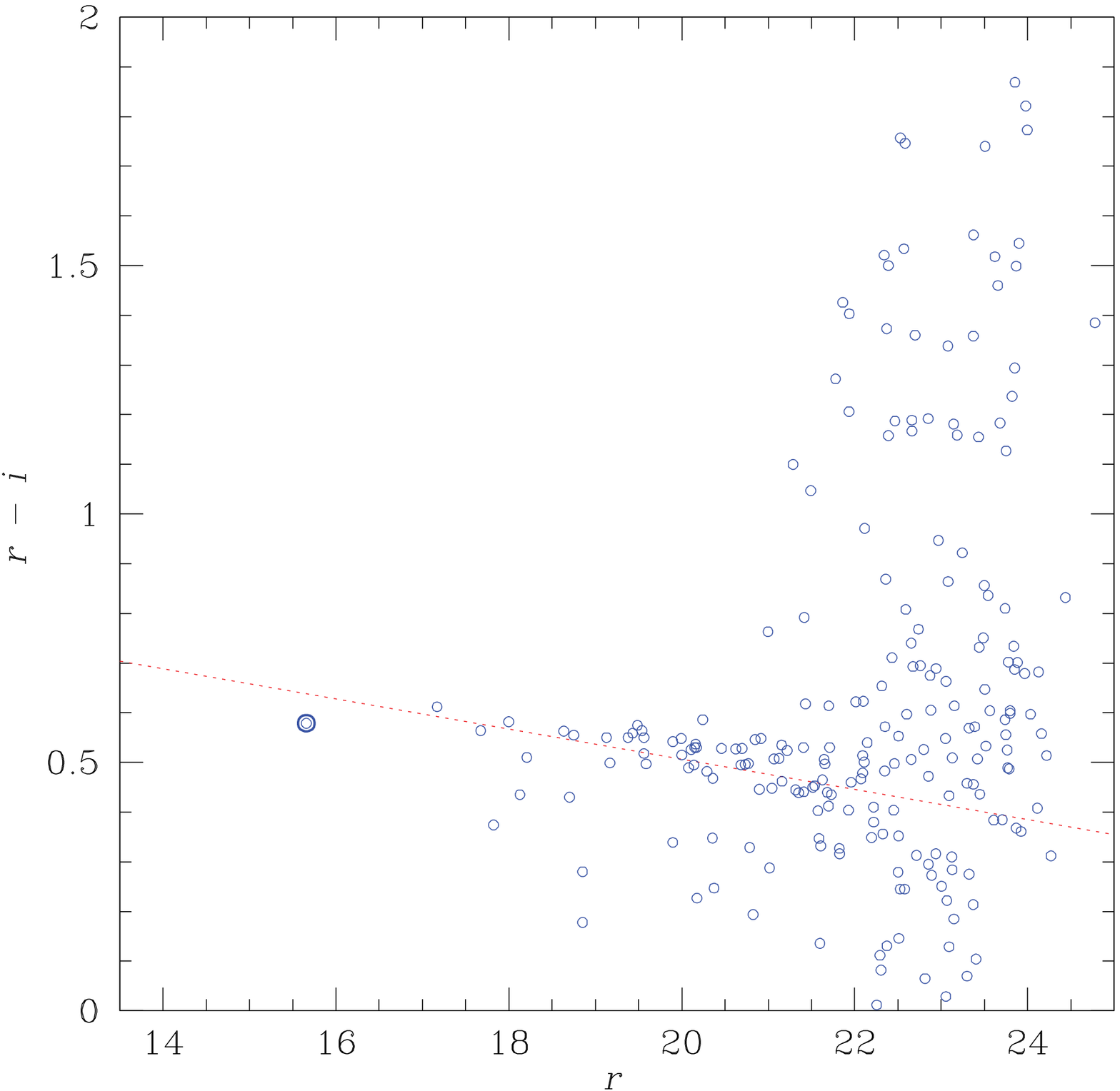}
  \medskip
  
  \caption{PSZ1 G076.44+23.53, a cluster that could be classified as a
    fossil group \citep[e.g.,][]{2010ApJ...708.1376V}.  {\it Left\/}:
    $i^\prime$-band RTT150 image.  {\it Right\/}: colour-magnitude
    diagram of extended objects near the cluster centre in the area
    enclosed by the dashed circle in the left panel.The dotted line
    indicates the red sequence.}
  \label{fig:p747}
  
  \vskip 1cm
  ~
\end{figure*}

\begin{figure}
  \centering
  \includegraphics[width=0.8\linewidth]{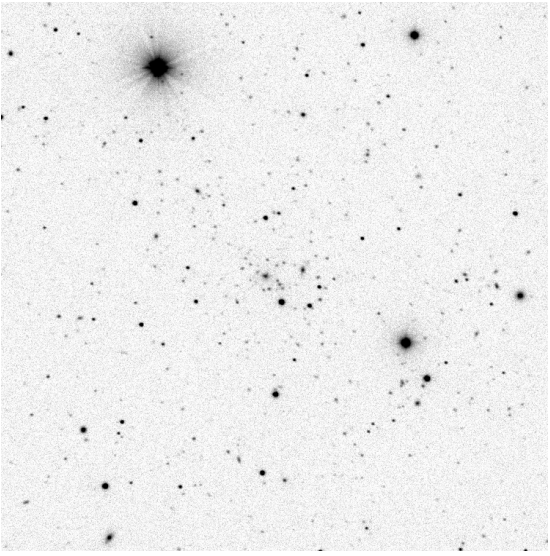}
  \medskip
  
  \caption{PSZ1 G048.22-65.03\quad RTT150 $r^\prime$-band 
    image. This cluster is too distant to be identified with DSS
    ($z\approx 0.42$) and there are no SDSS data in this field.}
  \label{fig:G04822}
\end{figure}

\begin{figure}
  \centering
  \includegraphics[width=0.8\linewidth]{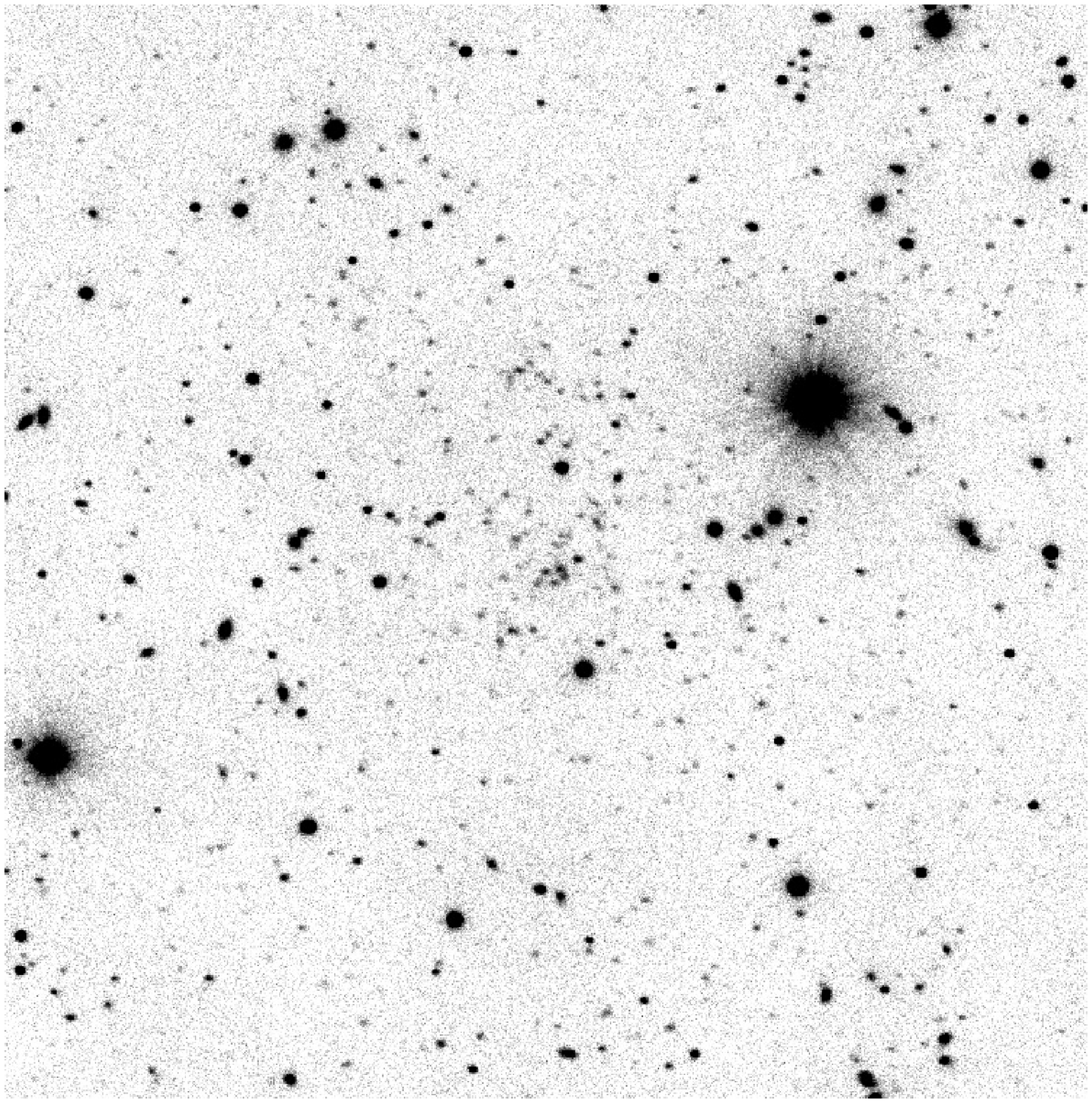}
  \medskip
  
  \caption{PSZ1 G084.04+58.75\quad RTT150  $i^\prime$-band 
    image. This cluster is too distant to be identified with SDSS
    ($z = 0.731$).}
  \label{fig:G08404}
\end{figure}

\subsection{Cluster identification examples}
\label{sec:ex}

\paragraph{PSZ1 G060.12+11.42:} Example of a cluster at low Galactic
latitude ($b\approx-11.4^\circ$). From Fig.~\ref{fig:g06012}, one can
see that it is difficult to identify the surface density enhancement
of cluster member galaxies here due to the large number of Galactic
stars in the field (left panel, $i^\prime$-band RTT150 image). This
cluster can be identified using the red sequence in the
colour-magnitude diagram (right panel).  It cannot be identified in
DSS, and there are no SDSS imaging data in this field.  Additional
imaging data were needed, easily obtained with a 1.5-m class
telescope.

\paragraph{PSZ1 G076.44+23.53:} Example of a cluster with a 
brightest cluster galaxy much more luminous than other cluster
galaxies. Clusters of this type are usually classified as fossil
groups \citep[e.g.,][]{2010ApJ...708.1376V}, see Fig.~\ref{fig:p747}.
This cluster appears as almost a single elliptical galaxy in DSS and
most of cluster member galaxies are not detected, since they are much
fainter than the brightest galaxy.  Such clusters are therefore
difficult to identify with DSS even at low redshifts, e.g., at
$z=0.169$ in this case.  There are no SDSS data in this field.
Additional imaging data were thus needed to identify this cluster.

\paragraph{PSZ1 G048.22-65.03:} This cluster is too distant to be
identified with DSS ($z\approx 0.42$, see Fig.~\ref{fig:G04822}). It
could have been identified at the depth of SDSS, but there are no SDSS
data for this field.

\paragraph{PSZ1 G084.04+58.75:} This cluster is too distant to be
identified with SDSS ($z = 0.731$, see Fig.~\ref{fig:G08404}). 


\subsection{Notes on individual objects}
\label{sec:notes}

\begin{figure}
  \centering
  \includegraphics[width=0.8\linewidth]{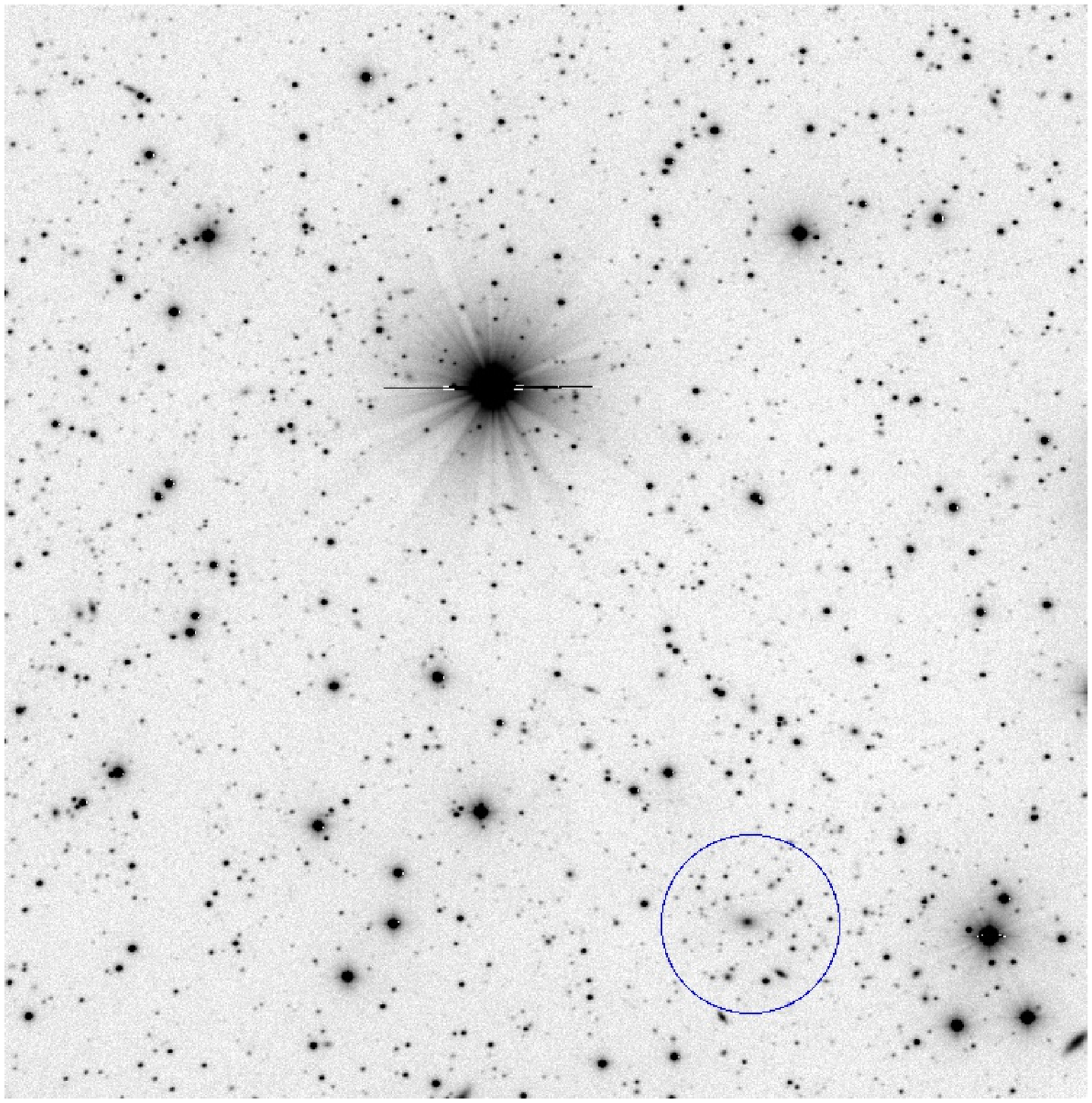}
  \medskip
  \caption{PSZ1 G066.01-23.30\quad  RTT150  $i^\prime$-band image centred on the \Planck\ SZ source coordinates. The position of the optical galaxy cluster is shown with a circle.}
  \label{fig:G066}
\end{figure}

\paragraph{PSZ1 G066.01-23.30:} There is a clear concentration of
galaxies that form a well-defined red sequence, shown within the
circle in Fig.~\ref{fig:G066}. However, the offset from the SZ source
centroid is large, about $4\arcmin$, so that there may be
astrophysical contamination or some other sources of SZ signal present
as well.  The spectrum of the central elliptical galaxy in this
concentration contains prominent emission lines. From the measured
intensity ratio $\lg(\hbox{[N{\sc II}]}\lambda6583/{\rm H}\alpha)
\approx 0$, we conclude that this galaxy contains a narrow-line AGN
\citep{veilleux87}.


\begin{figure}
  \centering
  \includegraphics[width=0.8\linewidth]{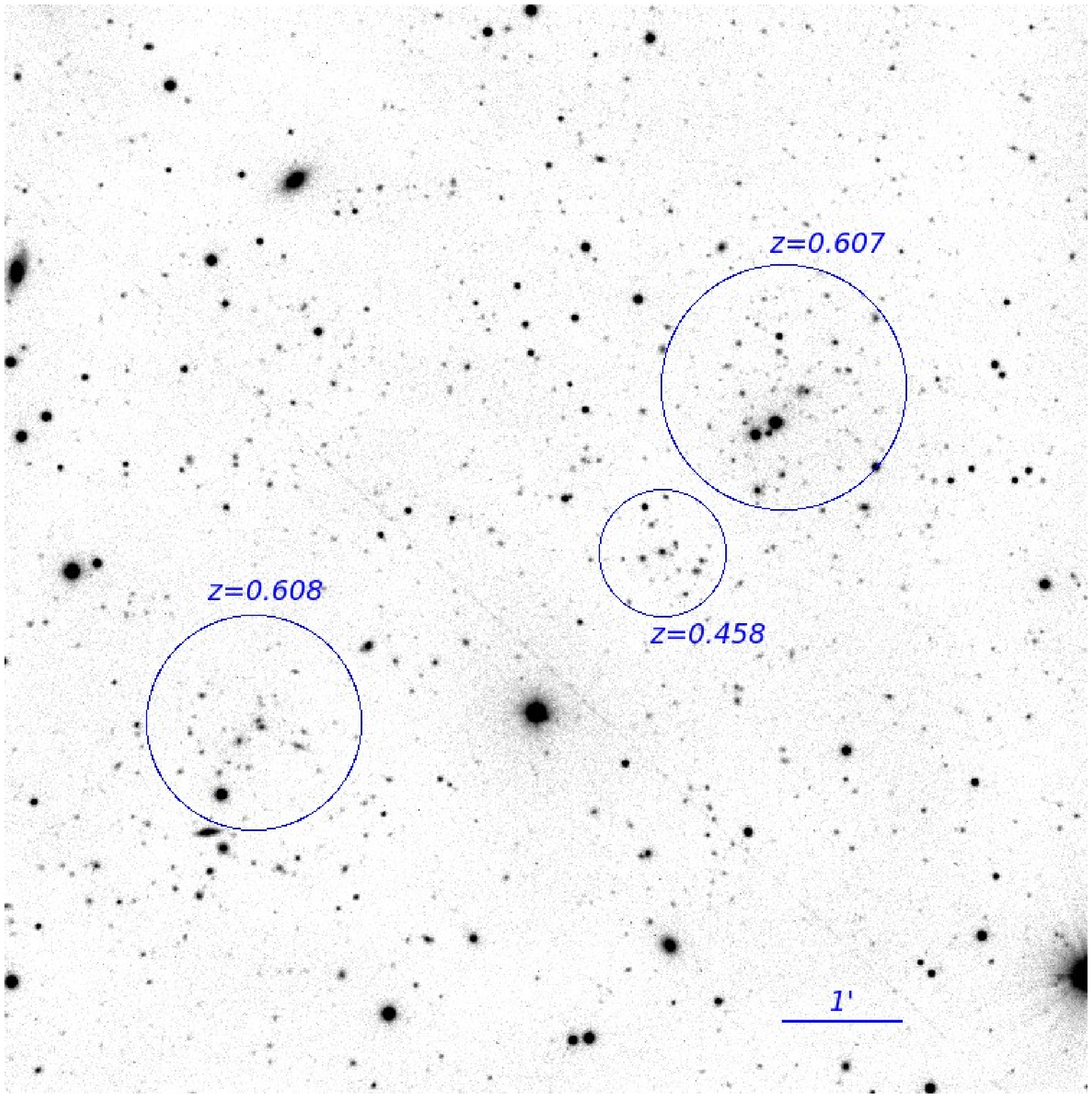}
  \medskip
  \caption{PSZ1 G070.91+49.26\quad  RTT150  $i^\prime$-band image centred on the \Planck\ SZ source coordinates. In addition to a double cluster at $z\approx0.61$, there is a smaller  foreground cluster at $z\approx0.46$.}
  \label{fig:G070}
\end{figure}

\paragraph{PSZ1 G070.91+49.26:} In addition to a double cluster at
$z=0.607$, there is also a smaller foreground cluster at
$z=0.458$ (the redshift was also measured at RTT), which should
produce an SZ signal.  The positions of these clusters in the field
centred on the \Planck\ SZ source coordinates are shown in
Fig.~\ref{fig:G070}. In the published PSZ1 catalogue this SZ source
was incorrectly identified with smaller cluster at $z=0.458$,
which should provide only a weak contribution to the measured SZ
signal as compared to the more distant and rich double cluster at
$z=0.607$.



\begin{figure}
  \centering
  \includegraphics[width=0.8\linewidth]{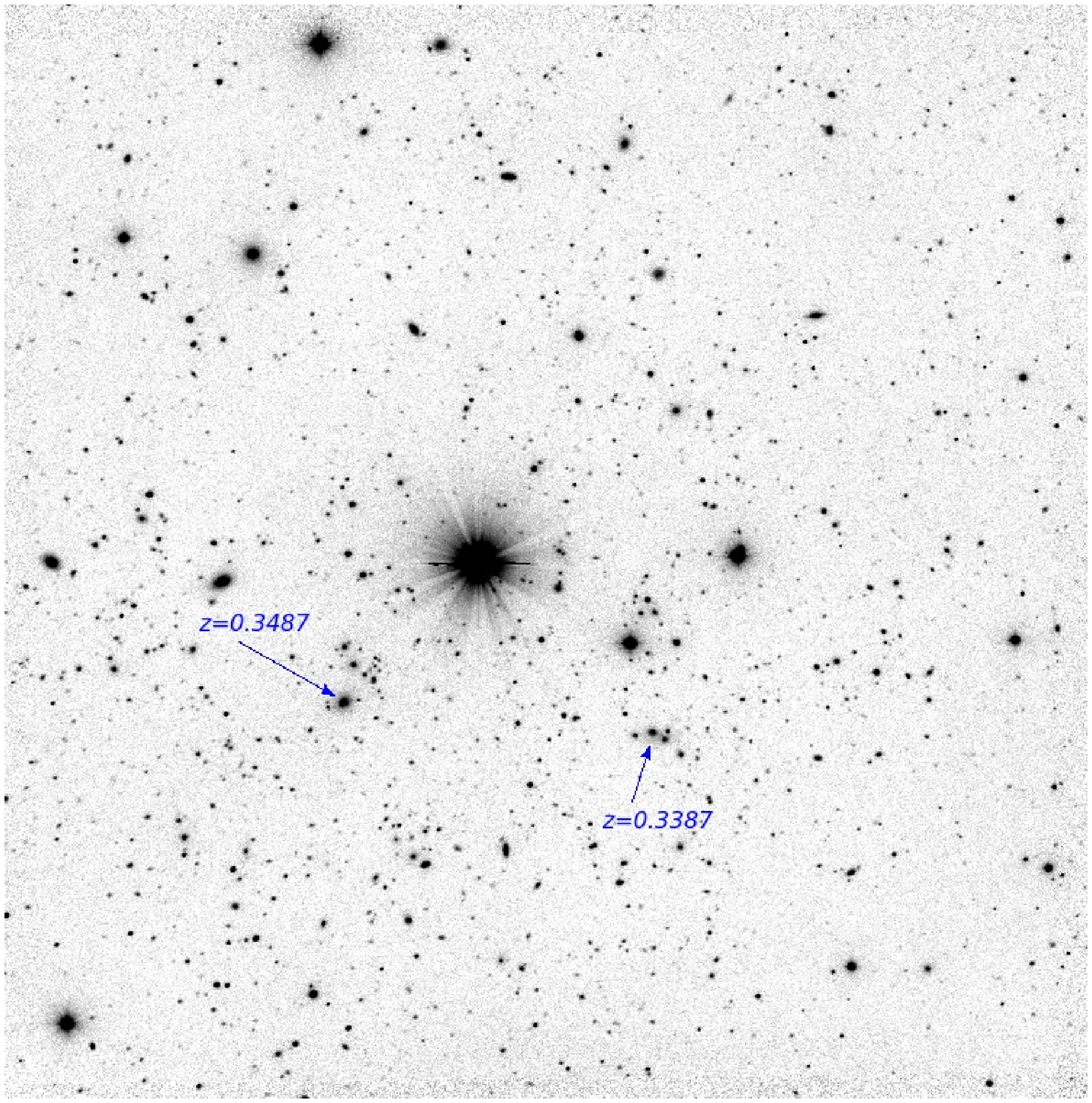}
  \medskip
  \caption{PSZ1 G092.27-55.73\quad  RTT150  $i^\prime$-band image.  Two cD
    galaxies and their redshifts are shown.}
  \label{fig:G092}
\end{figure}

\paragraph{PSZ1 G092.27-55.73:} While this object was photometrically
identified as a double cluster, redshift measurements show that the
two cD galaxies separated by $3\parcm4$ (Fig.~\ref{fig:G092})
are located at significantly different redshifts, $z=0.3487$ and
$z=0.3387$. The redshift difference is probably too large for a single
gravitationally bound object, therefore, this SZ detection likely
consists of two projected clusters. It is impossible to separate the
members of these two clusters in a photometric red sequence due to
their close redshifts. To estimate the richness of each cluster,
additional X-ray observations are required or, even better, spectroscopic
measurements.

\begin{figure}
  \centering
  \includegraphics[width=0.8\linewidth]{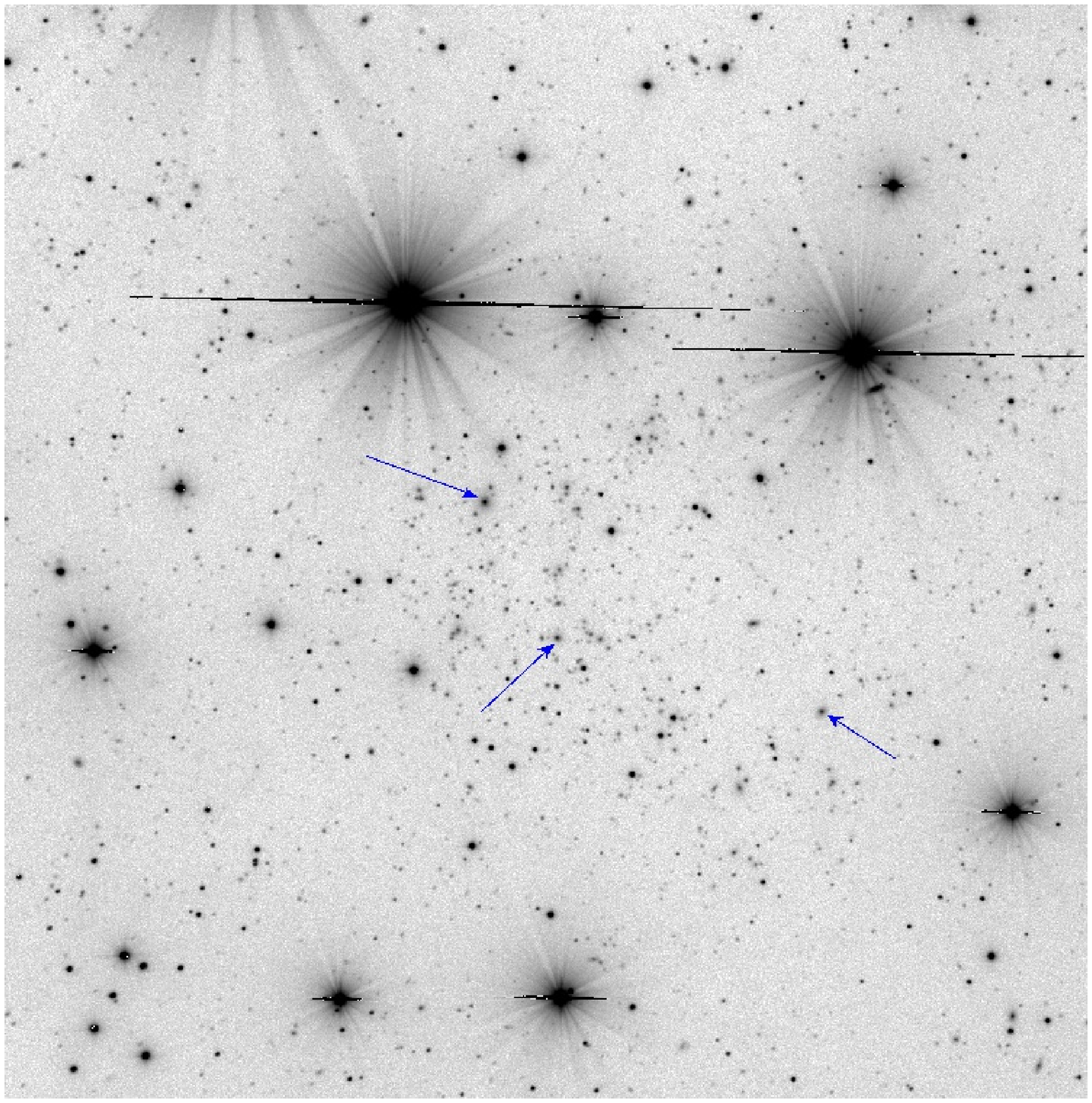}
  \medskip
  \caption{PSZ1 G100.18-29.68\quad RTT150 $i^\prime$-band image.
    The arrows indicate foreground elliptical galaxies.}
  \label{fig:G100}
\end{figure}

\paragraph{PSZ1 G100.18-29.68:} In addition to a rich galaxy cluster
at $z=0.485$, there are also a few foreground elliptical galaxies at
$z=0.178$, indicated with the arrows in
Fig.~\ref{fig:G100}.  These galaxies can also be identified in the
upper right panel of Fig.~\ref{fig:gri} as ones with bluer colours than the
member galaxies of the present cluster.

\begin{figure}
  \centering
  \includegraphics[width=0.8\linewidth]{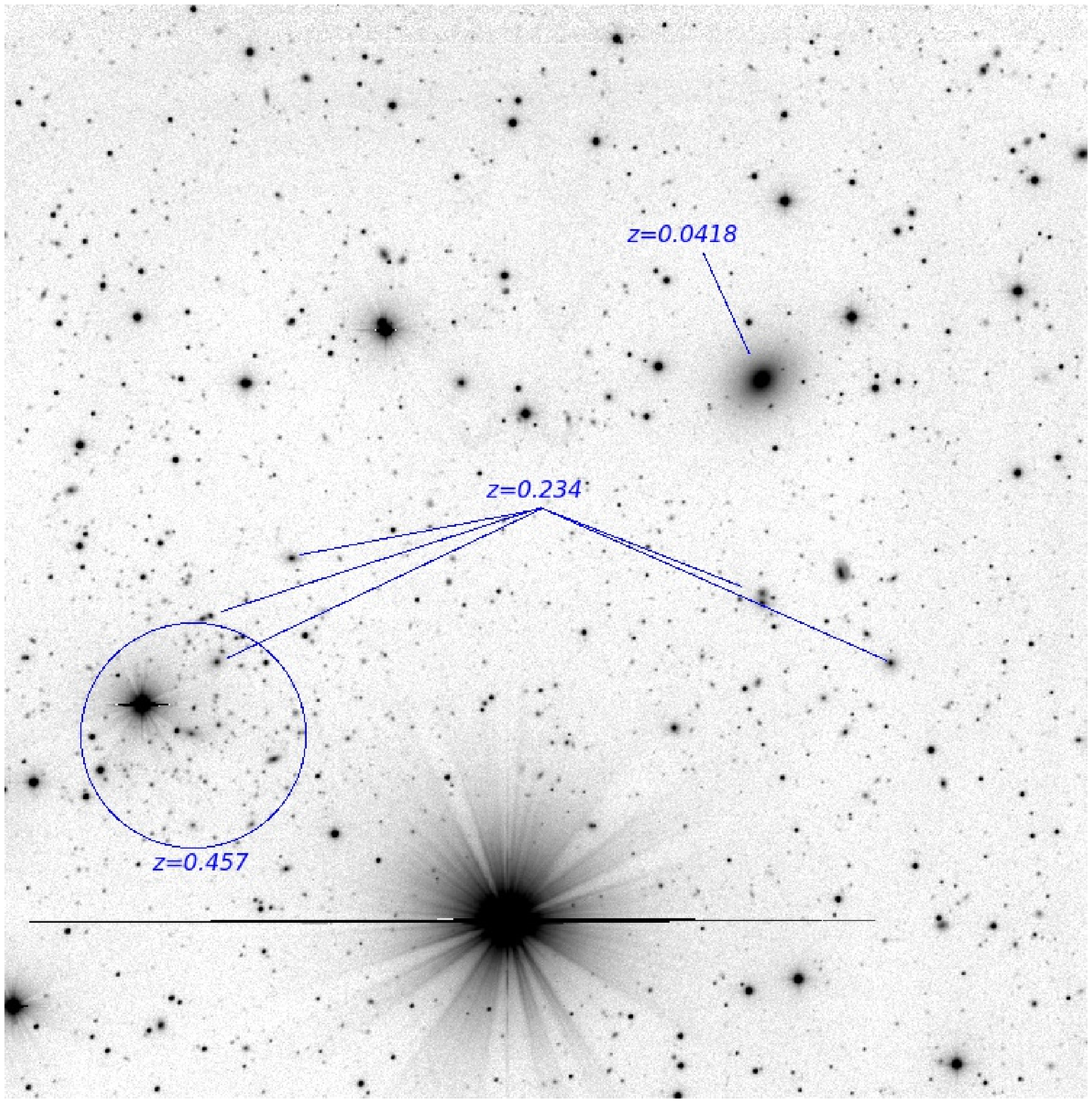}
  \medskip
  \caption{PSZ1 G109.14-28.02\quad  RTT150  $i^\prime$-band image. In
    addition to the rich cluster at $z=0.457$, there are also two
    galaxy groups at $z=0.234$, a nearby elliptical galaxy at
    $z=0.0418$, and a bright star in the field.}
  \label{fig:G109}
\end{figure}

\paragraph{PSZ1 G109.14-28.02:} In addition to the rich galaxy 
cluster at $z=0.457$, there are two galaxy groups at $z=0.234$ (the
redshift of one of these groups was also measured spectroscopically at
RTT150), a nearby elliptical galaxy at $z=0.0418$ (2MASX
J23524477+3319474), and a bright star in the field
(Fig.~\ref{fig:G109}). The main part of the SZ signal detected by
\Planck\ is most probably produced by the rich cluster at $z=0.457$;
however, other objects may also affect the photometry of the detected
SZ source.

\paragraph{PSZ1 G227.89+36.58:} This cluster is at $z\approx0.47$.  
There is also a foreground group offset by about 4\arcm\ to the SW at
$z=0.2845$ (ZwCl 0924.4+0511, MaxBCG J141.76983+04.97937). From
comparison of optical richnesses, we conclude that the cluster at
$z\approx0.47$ is much more massive, and should produce most of the SZ
signal detected by \Planck. The SZ photometry, however, may be
affected by the foreground group.

\subsection{Complicated cases}
\label{sec:complicated}

In a few cases, optical data are not sufficient to reliably identify
observed SZ sources. In order to determine the nature of these
objects, additional SZ or X-ray data are needed.

\begin{figure}
  \centering
  \includegraphics[width=0.8\linewidth]{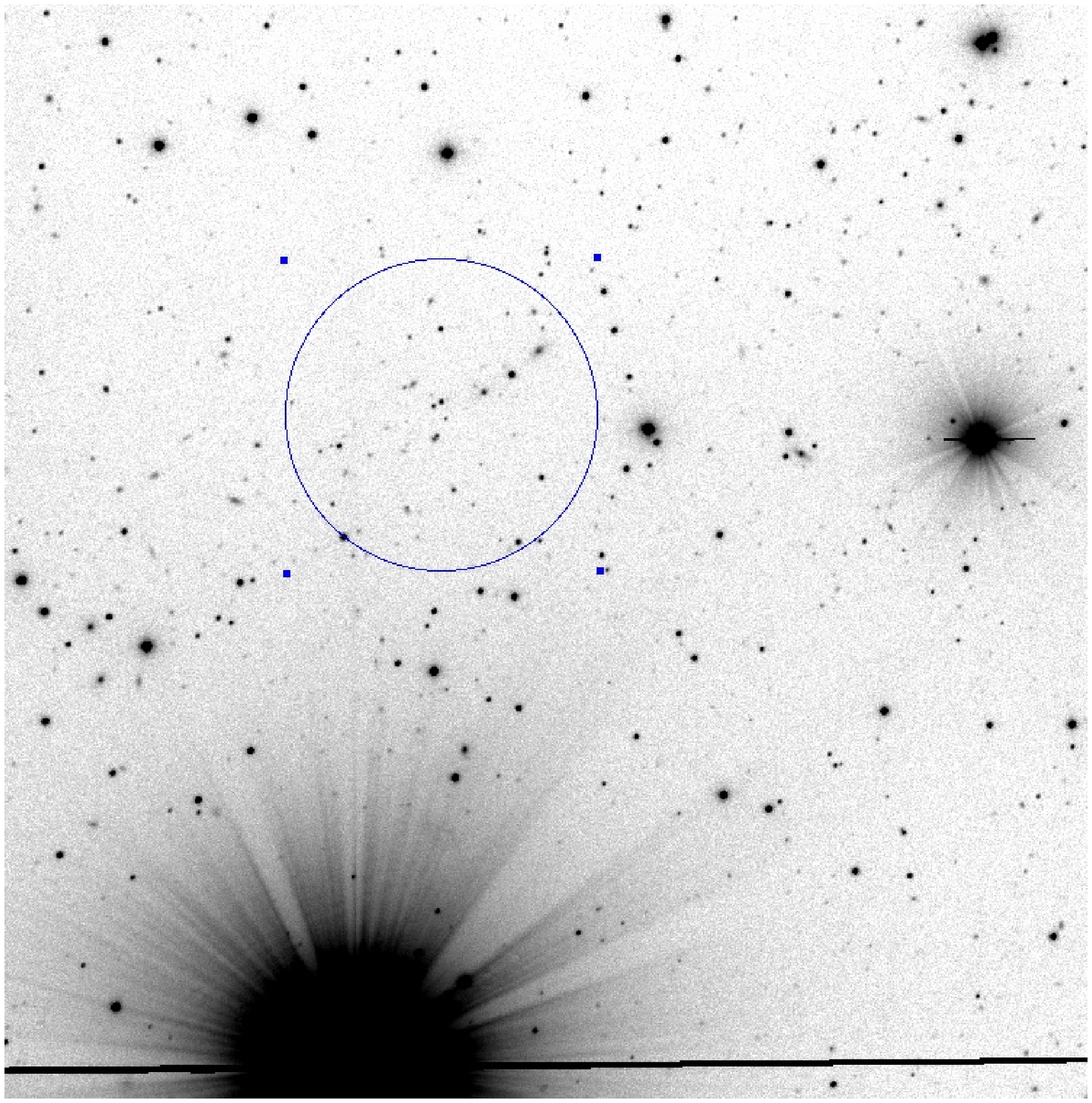}
  \medskip
  \caption{PSZ1 G103.50+31.36\quad RTT150 $i^\prime$-band image
    centred on the coordinates of the SZ source detected by \Planck,
    identified with the galaxy group shown in the circle. However,
    there is also a very bright star ($m_V=5.8$) at the edge of the
    field, 5\parcm6 from the \Planck\ position and detected by
    \Planck, which most probably affects the SZ signal.}
  \label{fig:G103}
\end{figure}

\paragraph{PSZ1 G103.50+31.36:} This SZ source may be identified 
with an irregular group of galaxies (Fig.~\ref{fig:G103}). Its
redshift can be estimated photometrically as $z\approx0.238$; however,
the very bright star HR\,6606 ($m_V=5.8$) is located at the edge of
the field, about 5\parcm6 from the \Planck\ position.  HR\,6606 is
detected by \Planck\ with flux density about 1\,Jy at 857\,GHz
\citep{PCCS13}, and probably affects the photometry of the detected SZ
source, artificially increasing its significance.

\begin{figure}
  \centering
  \includegraphics[width=0.8\linewidth]{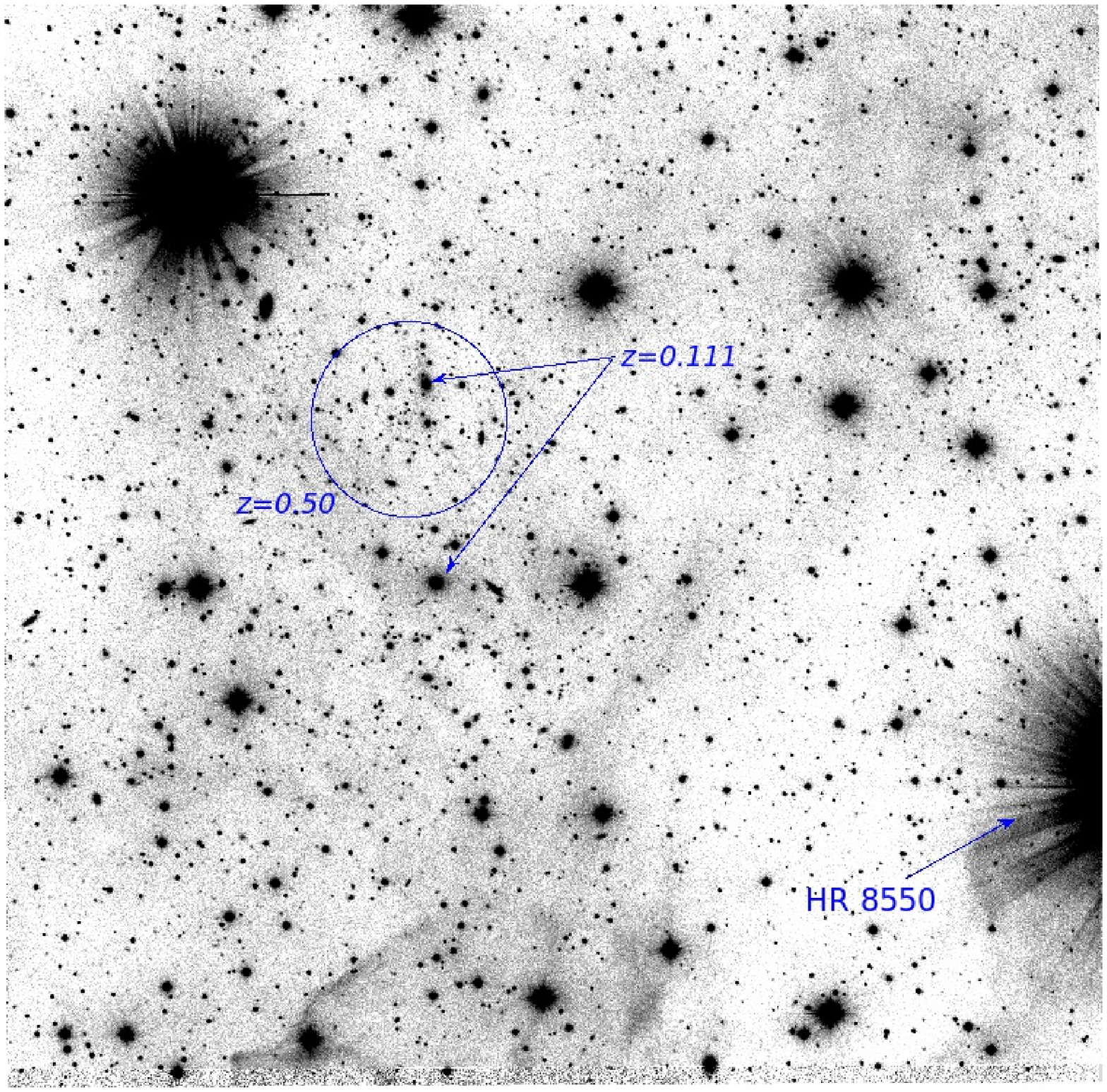}
  \medskip
  \caption{PSZ1 G115.70+17.51\quad RTT150 $i^\prime$-band image,
    centred on the coordinates of the SZ source detected by \Planck,
    which may be identified with the galaxy group shown in the circle.
    However, there are Galactic cirrus clouds throughout the field,
    and a very bright star ($m_V=6.8$) about 7\arcm\ away, which most
    probably affect the SZ signal.}
  \label{fig:G115}
\end{figure}

\paragraph{PSZ1 G115.70+17.51:} There is a group of galaxies with 
red sequence colour corresponding to a redshift of $z\approx0.50$ near
the centroid of the SZ detection (Fig.~\ref{fig:G115}), and also a few
galaxies at $z=0.1112$ (measured spectroscopically at RTT).  Hovewer,
there is a lot of Galactic cirrus in the field, as well as a very
bright star (HR\,8550, $m_V=6.8$) approximately 7\arcm\ from the SZ
position, which is detected by \Planck\ at 857\,GHz (where the
expected SZ increment is negligible) at about 10\,Jy \citep{PCCS13}.
Both the cirrus and the star likely affect the SZ signal detected by
{\it Planck}.

\medskip

We obtained deep $r^\prime$ and $i^\prime$ images with the RTT150 for
nine more objects from the PSZ1 cluster catalogue (G037.67+15.71,
G115.34-54.89, G115.59-44.47, G146.00-49.42, G159.26+71.11,
G167.43-38.04, G184.50-55.73, G194.68-49.73, and G199.70+37.01) and
for 30 sources from intermediate versions of the \Planck\ catalogue
that were eventually not confirmed as part of the PSZ1 catalogue.  Out
of the nine sources from the PSZ1 catalogue, in four cases
(G037.67+15.71, G159.26+71.11, G184.50-55.73, and G199.70+37.01) we
detected a few elliptical galaxies in the field, which, however, cannot
be identified as a galaxy cluster on the basis of our data.  In the
other cases, bright stars (G115.34-54.89 and G146.00-49.42) and
Galactic cirrus (G115.59-44.47, G167.43-38.04, and G146.00-49.42) most
probably affect the measurements. We conservatively maintained these
objects in the PSZ1 catalogue; however, they will need to be assessed
in the future by additional observations, in particular in X-rays.

\section{Conclusions}

This article is a companion paper to the \Planck\ catalogue of SZ
sources published in \cite{PSZcat13}. We present here the results of
approximately three years of optical observations of \Planck\ SZ
sources with the Russian-Turkish 1.5-m telescope. Approximately 60
clear dark and grey nights --- 20\,\% of the total clear dark and grey
time at the telescope during that period --- were used for these
observations. We also used approximately 32 hours of clear weather at
the BTA~6-m telescope of the SAO RAS.

In total, deep direct images were obtained of more than one hundred
fields in multiple filters.  We identified 47 previously unknown
clusters, 41 of which were included in the PSZ1 catalogue, and
selected galaxies to be used in determining cluster redshifts. We
measured redshifts of 65 \Planck\ clusters, including the redshifts of
12 distant clusters measured at the 6-m BTA telescope. Thirty-one of
these redshifts were measured after publication of the PSZ1 catalogue
\citep{PSZcat13} and are published here for the first time. For 14
more clusters, we give photometric redshift estimates. Some clusters
with only a few elliptical galaxies or with possible contamination by
stars and galactic dust have been kept in the PSZ1 catalogue.

We identified six cases of projections (see Sect.~\ref{sec:notes})
among the clusters identified in our work.  A similar result was
obtained earlier with a sample of \Planck\ clusters validated using
\emph{XMM}-{\it Newton} X-ray observations \citep{planck2012-IV}. The
fraction of projected clusters seems to be higher than in other
surveys. For example, only six out of about 200 clusters, either in
the \emph{400d} X-ray galaxy cluster survey \citep{400d} or in the
South Pole Telescope SZ survey \citep{sptclcat13}, are detected at
$<10\arcmin$ angular separation.

We emphasize that the sub-sample of \Planck\ SZ sources studied in our
work is not statistically representative in any sense.  The sources
observed at the RTT150 were selected from different versions of the
\Planck\ SZ source catalogue during a two-year period.  It is
therefore impossible to quantify biases in the selection of sources
from the \Planck\ catalogue for follow-up in this programme.  If a
high fraction of projections is determined in a statistically
representative sample of \Planck\ clusters, it would imply that the
detection probability of projected clusters is significantly enhanced
in the \Planck\ data, and that projection effects should be taken into
account in statistical calibration of the \Planck\ catalogue.

In particular, it is not possible to estimate a fraction of false
sources in the \Planck\ SZ survey from the RTT150 sample only.  We
might expect this fraction to be larger in the RTT150 sample than in
the full PSZ1 sample, since a large number of ``good'' clusters
immediately identified in the PSZ1 sample using DSS and SDSS data were
excluded from the RTT150 sample. But, on the other hand, targets for
observations were selected to confirm candidates suspected of being
clusters on the basis of other optical and IR data.

The RTT150 images typically reach $m_{i^\prime} = 23$\,mag.  Our study
shows that imaging data from a 1.5-m class telescope can be used
successfully to identify clusters below the SDSS limit ($z\approx0.6$)
at redshifts up to $z\approx1$, and also in fields where no SDSS
imaging data exist, e.g., at low Galactic latitudes in the North.  A
negative result in our optical identification programme does not
necessarily mean that an SZ detection is false.  Yet more distant
clusters may be identified using better optical data and also with
better data in the SZ and X-ray domains.  Follow-up programmes are now
also runming at other telescopes --- NOT, INT, GTC, TNG, WHT, NTT, and
others --- and optical identifications of \Planck\ cluster candidates
could be completed within a few years.

\begin{acknowledgements}

  The development of \Planck\ has been supported by: ESA; CNES and
  CNRS/INSU-IN2P3-INP (France); ASI, CNR, and INAF (Italy); NASA and
  DoE (USA); STFC and UKSA (UK); CSIC, MICINN, JA, and RES (Spain);
  Tekes, AoF, and CSC (Finland); DLR and MPG (Germany); CSA (Canada);
  DTU Space (Denmark); SER/SSO (Switzerland); RCN (Norway); SFI
  (Ireland); FCT/MCTES (Portugal); and PRACE (EU).  The authors thank
  TUBITAK, IKI, KFU, and AST for support in using the RTT150
  (Russian-Turkish 1.5-m telescope, Bakyrlytepe, Turkey), and in
  particular thank KFU and IKI for providing a significant amount of
  their observing time.  We also thank the BTA 6-m telescope Time
  Allocation Committee (TAC) for support of the optical follow-up
  project. We are grateful to S.~N.~Dodonov, A.~Galeev, E.~Irtuganov,
  S.~Melnikov, A.~V.~Mescheryakov, A.~Moiseev, A.~Yu.~Tkachenko,
  R.~Uklein, R.~Zhuchkov and to other observers at the RTT150 and BTA
  6-m telescopes for their help with the observations. This work was
  supported by Russian Foundation for Basic Research, grants
  13-02-12250-ofi-m, 13-02-01464, 14-22-03111-ofi-m, by Programs of
  the Russian Academy of Sciences P-21 and OPhN-17 and by the subsidy
  of the Russian Government to Kazan Federal University (agreement
  No.02.A03.21.0002). This research has made use of the following
  databases: the NED database, operated by the Jet Propulsion
  Laboratory, California Institute of Technology, under contract with
  NASA; SIMBAD, operated at CDS, Strasbourg, France; and the SZ
  database operated by Integrated Data and Operation Center (IDOC)
  operated by IAS under contract with CNES and CNRS.

\end{acknowledgements}

\bibliographystyle{aa}

\end{document}